\documentclass[english,longbibliography,nofootinbib,superscriptaddress,12pt,sort&compress,showkeys]{revtex4}
\usepackage[T1]{fontenc}
\usepackage[latin9]{inputenc}
\usepackage{float}
\usepackage{booktabs}
\usepackage{amsmath}
\usepackage{graphicx}

\makeatletter

\providecommand{\tabularnewline}{\\}
\newcommand{\lyxdot}{.}

\@ifundefined{textcolor}{}
{%
 \definecolor{BLACK}{gray}{0}
 \definecolor{WHITE}{gray}{1}
 \definecolor{RED}{rgb}{1,0,0}
 \definecolor{GREEN}{rgb}{0,1,0}
 \definecolor{BLUE}{rgb}{0,0,1}
 \definecolor{CYAN}{cmyk}{1,0,0,0}
 \definecolor{MAGENTA}{cmyk}{0,1,0,0}
 \definecolor{YELLOW}{cmyk}{0,0,1,0}
}

\usepackage{indentfirst}
\usepackage{amsfonts}
\usepackage[T1]{fontenc}
\usepackage{ae,aecompl}
\usepackage{sidecap}
\usepackage[section]{placeins}
\usepackage{epsf}
\makeatother

\usepackage{babel}
\date{\today}

\makeatother

\usepackage{babel}
\begin{document}
\title{Development of a physically-informed neural network interatomic potential
for tantalum}
\author{Yi-Shen Lin, Ganga P. Purja Pun and Yuri Mishin}
\address{Department of Physics and Astronomy, MSN 3F3, George Mason University,
Fairfax, Virginia 22030, USA}
\begin{abstract}
Large-scale atomistic simulations of materials heavily rely on interatomic
potentials, which predict the system energy and atomic forces. One
of the recent developments in the field is constructing interatomic
potentials by machine-learning (ML) methods. ML potentials predict
the energy and forces by numerical interpolation using a large reference
database generated by quantum-mechanical calculations. While high
accuracy of interpolation can be achieved, extrapolation to unknown
atomic environments is unpredictable. The recently proposed physically-informed
neural network (PINN) model significantly improves the transferability
by combining a neural network regression with a physics-based bond-order
interatomic potential. Here, we demonstrate that general-purpose PINN
potentials can be developed for body-centered cubic (BCC) metals.
The proposed PINN potential for tantalum reproduces the reference
energies within 2.8 meV/atom. It accurately predicts a broad spectrum
of physical properties of Ta, including (but not limited to) lattice
dynamics, thermal expansion, energies of point and extended defects,
the dislocation core structure and the Peierls barrier, the melting
temperature, the structure of liquid Ta, and the liquid surface tension.
The potential enables large-scale simulations of physical and mechanical
behavior of Ta with nearly first-principles accuracy while being orders
of magnitude faster. This approach can be readily extended to other
BCC metals.
\end{abstract}
\maketitle

\section*{Introduction}

The critical ingredient of all large-scale molecular dynamics (MD)
and Monte Carlo (MC) simulations of materials is the classical interatomic
potential, which predicts the system energy and atomic forces as a
function of atomic positions and, for multicomponent systems, their
occupation by chemical species. Computations with interatomic potentials
are much faster than quantum-mechanical calculations explicitly treating
the electrons. The computational efficiency of interatomic potentials
enables simulations on length scales up to $\sim10^{2}$ nm ($\sim10^{7}$
atoms) and time scales up to $\sim10^{2}$ ns. 

Interatomic potentials partition the total potential energy $E$ into
a sum of energies $E_{i}$ assigned to individual atoms $i$: $E=\sum_{i}E_{i}$.
Each atomic energy $E_{i}$ is expressed as a function of the local
atomic positions $\mathbf{R}_{i}\equiv(\mathbf{r}_{i1},\mathbf{r}_{i2},...,\mathbf{r}_{in_{i}})$
in the vicinity of the atom. The form of the potential function
\begin{equation}
E_{i}=\Phi(\mathbf{R}_{i},\mathbf{p})\label{eq:1}
\end{equation}
must ensure the invariance of energy under rotations and translations
of the coordinate axes, and permutations of the atoms. The partitioning
into atomic energies makes the total energy computation a linear-$N$
procedure ($N$ being the number of atoms), enabling effective parallelization
by domain decomposition. Physically, this partitioning is only justified
for systems with short-range interactions. The potential function
$\Phi$ in Eq.(\ref{eq:1}) additionally depends on a set of adjustable
parameters $\mathbf{p}=(p_{1},...,p_{m})$, which are optimized by
training on a reference database. Once the optimization is complete,
the parameters are fixed once and for all and used in all subsequent
simulations. The atomic forces required for MD simulations are obtained
by differentiation of the total energy with respect to atomic coordinates.

Potentials can be divided into two categories according to their intended
usage. General-purpose potentials are trained to reproduce a broad
spectrum of properties that are most essential for atomistic simulations.
The reference structures must be diverse enough to represent the most
typical atomic environments occurring in simulations. Once published,
the potential is used for almost any type of simulation of the material.
Special-purpose potentials are designed for one particular type of
simulation and are not expected to be transferable to other applications. 

Two major classes of potentials are the traditional potentials and
the emerging class of machine-learning (ML) potentials \citep{Behler:2016aa,Botu:2017aa,Deringer:2019aa,Zuo:2020aa}.
The traditional potentials use a functional form of $\Phi(\mathbf{R}_{i},\mathbf{p})$
in Eq.(\ref{eq:1}) that reflects the basic physics of interatomic
bonding in the given material. Accordingly, such functional forms
are specific to particular classes of materials. For example, the
embedded atom method (EAM) \citep{Daw83,Daw84,Finnis84}, the modified
EAM (MEAM) \citep{Baskes87}, and the angular-dependent potential
(ADP) \citep{Mishin05a} are designed for metallic systems. The Tersoff
\citep{Tersoff88,Tersoff:1988dn,Tersoff:1989wj} and Stillinger-Weber
\citep{Stillinger85} potentials were specifically developed for strongly
covalent materials such as silicon and carbon. Traditional potentials
depend on a small ($\sim10$) number of parameters, which are trained
on a small database of experimental properties and first-principles
energies or forces. The accuracy of traditional potentials is limited
compared to both the first-principles calculations and the ML potentials.
However, due to the physical underpinnings, traditional potentials
often demonstrate reasonable transferability to atomic configurations
lying well outside the training dataset. As long as the nature of
chemical bonding remains the same as assumed during the potential
construction, the predicted energies and forces may not be very accurate
but at least remain physically meaningful. Most of the traditional
potentials are general-purpose type.

The ML potentials are based on a different philosophy. The physics
of interatomic bonding is not considered beyond the principle of locality
of atomic interactions and the invariance of energy. The potential
function (\ref{eq:1}) is a high-dimensional nonlinear regression
implemented numerically. This function depends on a large ($\sim10^{3}$)
number of parameters, which are trained on a database containing $10^{3}$
to $10^{4}$ supercells whose energies or forces (or both) are obtained
by high-throughput density functional theory (DFT) \citep{hohenberg64:dft,kohn65:inhom_elec}
calculations. An ML potential computes the energy in two steps. First,
the local position vector $\mathbf{R}_{i}$ is mapped onto a set of
local structural parameters $\mathbf{G}_{i}=(G_{i1},G_{i2},...,G_{iK})$,
which encode the local environment and are invariant under rotations,
translations, and relabeling of atoms. Behler and Parrinello \citep{Behler:2007aa}
proposed that the size $K$ of the descriptors $\mathbf{G}_{i}$ be
the same for all atoms, even though the number of neighbors $n_{i}$
can vary from one atom to another. At the second step, the $K$-dimensional
descriptor space is mapped onto the 1D space of atomic energies. This
mapping is implemented by a pre-trained nonlinear regression $\mathcal{R}$.
Thus, the atomic energy calculation can be represented by the formula
$\mathbf{R}_{i}\rightarrow\mathbf{G}_{i}\overset{\mathcal{R}}{\rightarrow}E_{i}$.
Several regression methods have been employed, such as the Gaussian
process regression \citep{Payne.HMM,Bartok:2010aa,Bartok:2013aa,Li:2015aa,Glielmo:2017aa,Bartok_2018,Deringer:2018aa},
the kernel ridge regression \citep{Botu:2015bb,Botu:2015aa,Mueller:2016aa},
artificial neural networks (NN) \citep{Behler07,Bholoa:2007aa,Behler:2008aa,Sanville08,Eshet2010,Handley:2010aa,Behler:2011aa,Behler:2011ab,Sosso2012,Behler:2015aa,Behler:2016aa,Schutt:148aa,Imbalzano:2018aa},
the spectral neighbor analysis \citep{Thompson:2015aa,Chen:2017ab,Li:2018aa},
and the moment tensor potentials \citep{Shapeev:2016aa}. 

The development of an ML potential is a complex high-dimensional problem,
which is solved by applying ML methods for the reference database
generation, training, and error quantification. With the large number
of parameters available, an ML potential can be trained to reproduce
the reference database within a few meV/atom, which is the intrinsic
uncertainty of DFT calculations. Since the potential format is independent
of the nature of chemical bonding, ML potentials are not specific
to any particular class of materials. The same procedure is applied
to develop a potential for a metal, nonmetal, or a mixed-bonding system.
However, the high accuracy and flexibility come at a price: ML potentials
are effective numerical interpolators but poor extrapolators. Since
the energy and force predictions are not guided by any physics, extrapolation
outside the domain of known environments is unpredictable and often
unphysical. Nearly all ML potentials are special-purpose type; development
of general-purpose ML potentials is an extremely challenging task
\citep{Bartok_2018,Pun:2020aa}. 

\begin{figure}[H]
\noindent \begin{centering}
\includegraphics[width=0.85\columnwidth]{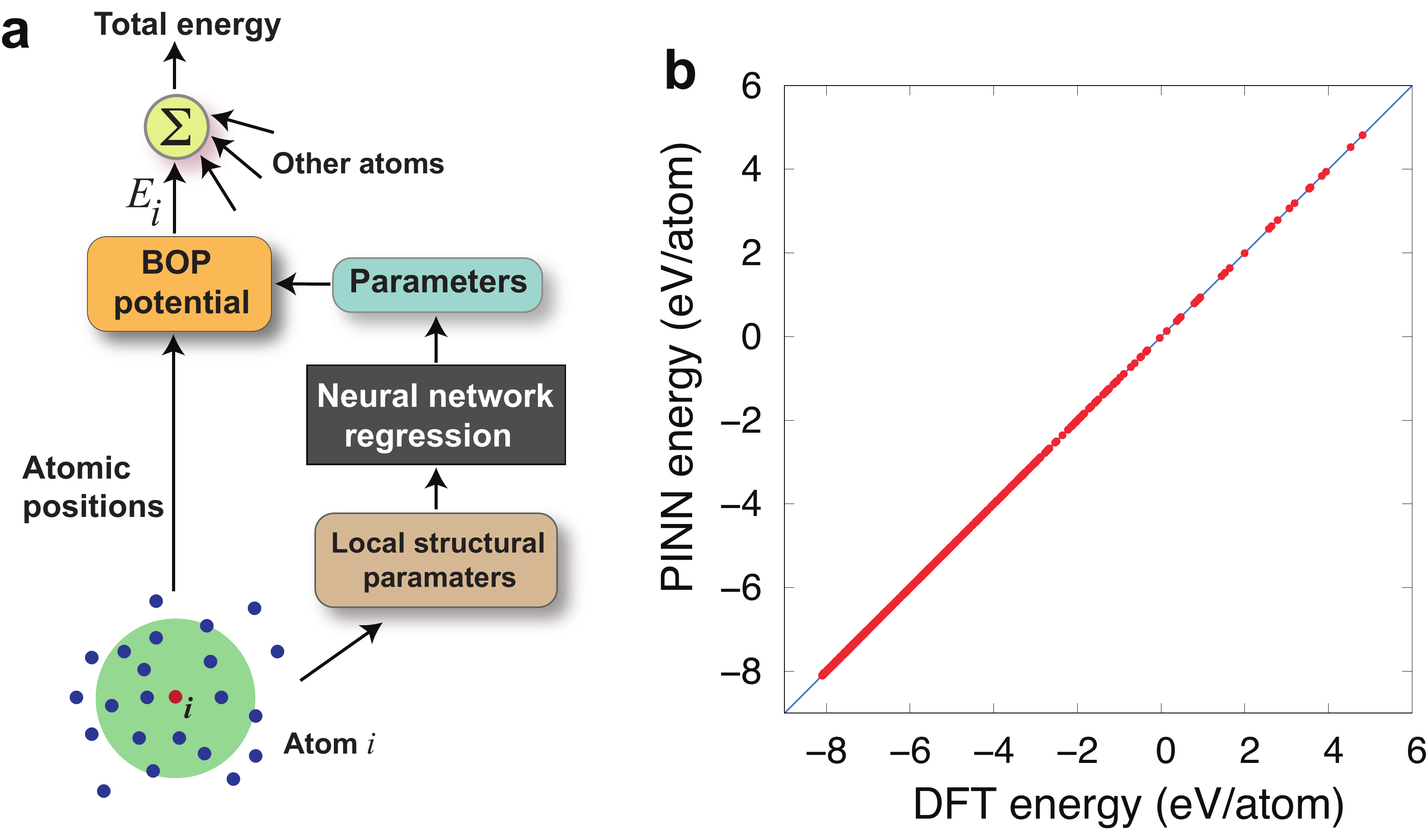}
\par\end{centering}
\caption{\label{fig:EneTraining}How the PINN potential works. \textbf{a} Flow
diagram of energy calculation in the PINN model. \textbf{b} Energies
of atomic configurations in the Ta training set computed with the
PINN potential versus the DFT energies (red dots). The blue line represents
the perfect fit.}
\end{figure}

The recently proposed physically-informed neural network (PINN) method
\citep{PINN-1} takes the best from both worlds by integrating an
ML regression with a physics-based interatomic potential. Instead
of directly predicting the atomic energy, the regression predicts
the best set of potential parameters $\mathbf{p}_{i}$ appropriate
to the given environment. The potential $\Phi(\mathbf{R}_{i},\mathbf{p}_{i})$
then computes the atomic energy using the predicted parameter set
$\mathbf{p}_{i}$. Thus, the formula of the method is (Fig.~\ref{fig:EneTraining}a)
\begin{equation}
\mathbf{R}_{i}\rightarrow\mathbf{G}_{i}\overset{\mathcal{R}}{\rightarrow}\mathbf{p}_{i}\overset{\Phi}{\rightarrow}E_{i}.\label{eq:2}
\end{equation}
The method drastically improves transferability to new environments
because the extrapolation is now guided by the physics embedded in
the interatomic potential rather than a purely mathematical algorithm.
This general idea can be realized with any regression method and any
meaningful interatomic potential. PINN is a particular realization
of this approach using a NN regression and an analytical bond-order
potential (BOP) \citep{PINN-1}. The latter has a functional form
general enough to work for both metals and nonmetals. Specifically,
the potential represents pairwise repulsions and attractions of the
atoms, the bond-order effect (bond weakening with the number of bonds),
the angular dependence of the bond energies, the screening of bonds
by surrounding atoms, and the promotion energy. The interactions extend
to neighbors within a 0.5 to 0.6 nm distance with a smooth cutoff.
More details about the BOP potential can be found in the Methods section. 

The original PINN formulation \citep{PINN-1} has been recently improved
by introducing a global BOP potential trained on the entire reference
database. Since the optimized parameter set ($\mathbf{p}^{0}$) is
small, the error of fitting is relatively large ($\sim10^{2}$ meV/atom).
A pre-trained NN then adds to $\mathbf{p}^{0}$ a set of local ``perturbations''
$\delta\mathbf{p}_{i}=(\delta p_{i1},...,\delta p_{im})$ to obtain
the final parameter set $\mathbf{p}_{i}=\mathbf{p}^{0}+\delta\mathbf{p}_{i}$.
The latter is used to predict the atomic energy $E_{i}=\Phi(\mathbf{R}_{i},\mathbf{p}^{0}+\delta\mathbf{p}_{i})$.
In this scheme, the energy predictions are largely guided by the global
BOP potential, which provides a smooth and physically meaningful extrapolation
outside the training domain. The magnitudes of the perturbations are
kept as small as possible. The same DFT level of accuracy is achieved
during the training as in the original PINN formulation \citep{PINN-1},
no computational overhead is incurred, but the transferability is
improved significantly. 

The modified PINN method has been recently applied to develop a general-purpose
ML potential for the face-centered cubic (FCC) Al \citep{Pun:2020aa}.
Here, we demonstrate that the method can also generate highly accurate
and transferable general-purpose potentials for body-centered cubic
(BCC) metals. We chose tantalum as a representative transition BCC
metal, in which the interatomic bonding has a significant directional
component due to the $d$-electrons. In addition to many structural
applications of Ta and Ta-based alloys \citep{Buckman:2000aa,Sungail:2020aa},
porous Ta is a promising biomaterial for orthopedic applications due
to its excellent biocompatibility and a favorable combination of physical
and mechanical properties \citep{Balla:2010aa}. This work paves the
way for the development of PINN potentials for other BCC metals in
the future.

\section*{Results}

\textbf{The potential development.} The reference database has been
generated by DFT calculations employing the Vienna Ab Initio Simulation
Package (VASP) \citep{Kresse1996,Kresse1999} (see the Methods section
for details). The database consists of the energies of 3,552 supercells
containing from 1 to 250 atoms. The supercells represent energy-volume
relations for 8 crystal structures of Ta, 5 uniform deformation paths
between pairs of structures, vacancies, interstitials, surfaces with
low-index orientations, 4 symmetrical tilt grain boundaries, $\gamma$-surfaces
on the (110) and (211) fault planes, a $\tfrac{1}{2}${[}111{]} screw
dislocation, liquid Ta, and several isolated clusters containing from
2 to 51 atoms. Some of the supercells contain static atomic configurations.
However, most are snapshots of \emph{ab initio} MD simulations at
different densities, and temperatures ranging from 293 K to 3300 K.
The BCC structure was sampled in the greatest detail, including a
wide range of isotropic and uniaxial deformations. The database represents
a total of 161,737 highly diverse atomic environments occurring in
typical atomistic simulations. More detailed information about the
database can be found in Supplementary Tables 1 and 2. About 90\%
of the supercells, representing 136,177 environments, were randomly
selected for training. The remaining 10\% were set aside for cross-validation. 

The potential training process involves several hyper-parameters describing
the NN architecture, the number and types of the local structural
parameters, and the regularization coefficients in the loss function.
We tested many combinations of these parameters before choosing the
final version. We emphasize that, for almost any choice of the hyper-parameters,
the potential could be trained to an error of about 3 meV/atom. However,
the predicted properties of Ta were different. Even different random
initializations of the NN's weights and biases resulted in slightly
different combinations of Ta properties. Hundreds of initialization-training
cycles with automated property testing had to be performed before
a potential that we deemed the best was selected. The hyper-parameters
and the properties reported below refer to the final version of the
potential.

As the local structural descriptors, we chose products of a radial
function and an angular function. The radial function is a Gaussian
peak of width $\sigma=1$ \AA{} centered at radius $r_{0}$ and smoothly
truncated at a cutoff $r_{c}$ = 4.8 \AA{} within the range $d$ =
1.5 \AA{} (see the Methods section). The angular part is a Legendre
polynomial $P_{l}(\cos\theta_{ijk})$, where $\theta_{ijk}$ is the
angle between the bonds $ij$ and $ik$. The size of the descriptors
$\mathbf{G}_{i}$ is $K=40$, corresponding to the set of Gaussian
positions $r_{0}=\{2.4,2.8,3.0,3.2,3.4,3.6,4.0,4.4\}$\,Å and the
Legendre polynomials of orders $l=0,1,2,4$ and $6$. 

The NN has a feed-forward $40\times32\times32\times8$ architecture
with 32 nodes in each of the hidden layers and 8 nodes in the output
layer. The latter corresponds to the $m=8$ perturbations to the BOP
parameters delivered by the NN. The total number of weights and biases
in the NN is 2,632. The loss function represents the mean-square deviation
of the predicted supercell energies from the DFT energies, augmented
by two regularization terms as detailed in the Methods section. The
NN's weights and biases were optimized by the L-BFGS unconstrained
minimization algorithm \citep{Fletcher:1987aa} to reach the root-mean-square
error (RMSE) of 2.80 meV/atom. Fig.~\ref{fig:EneTraining}b demonstrates
the accurate and unbiased agreement between the PINN and DFT energies
over a 13 eV/atom wide energy interval. Although atomic forces were
not used during the training, the forces predicted by the potential
were compared with DFT forces after the training. A strong correlation
is observed (Supplementary Fig.~1) with the RMSE of about 0.3 eV/\AA .
Note that the comparison includes forces as large as 20 eV/\AA .

10-fold cross-validation was performed to verify that the database
was not overfitted. At each rotation, the set-aside dataset mentioned
above replaced a similar number of supercells in the training set.
The training was repeated, and the RMSE of the potential was computed
on the validation dataset unseen during the training. The RMSE of
validation averaged over the 10 rotations is 2.89 meV/atom.

\medskip{}

\begin{figure}
\noindent \begin{centering}
\includegraphics[width=0.6\columnwidth]{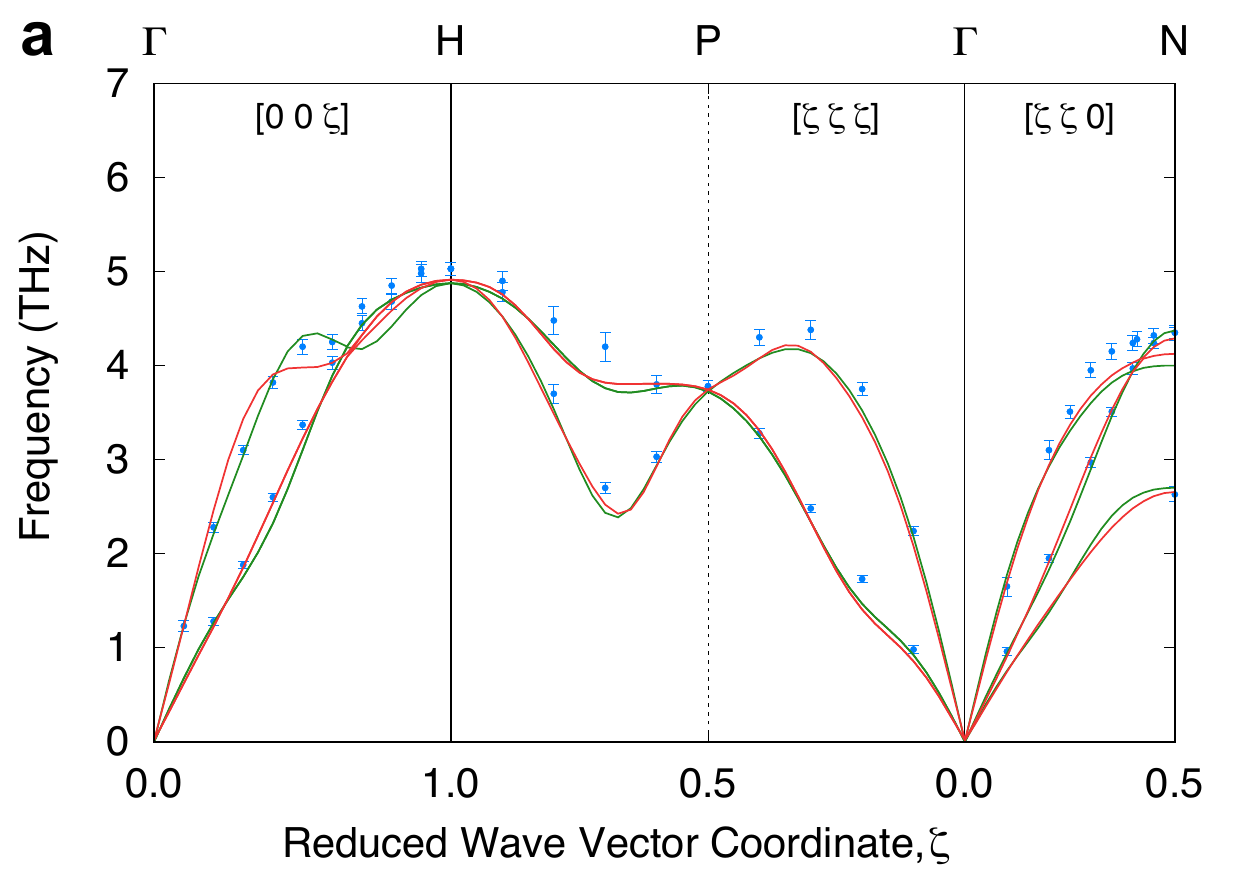}
\par\end{centering}
\bigskip{}

\noindent \begin{centering}
\includegraphics[width=0.6\textwidth]{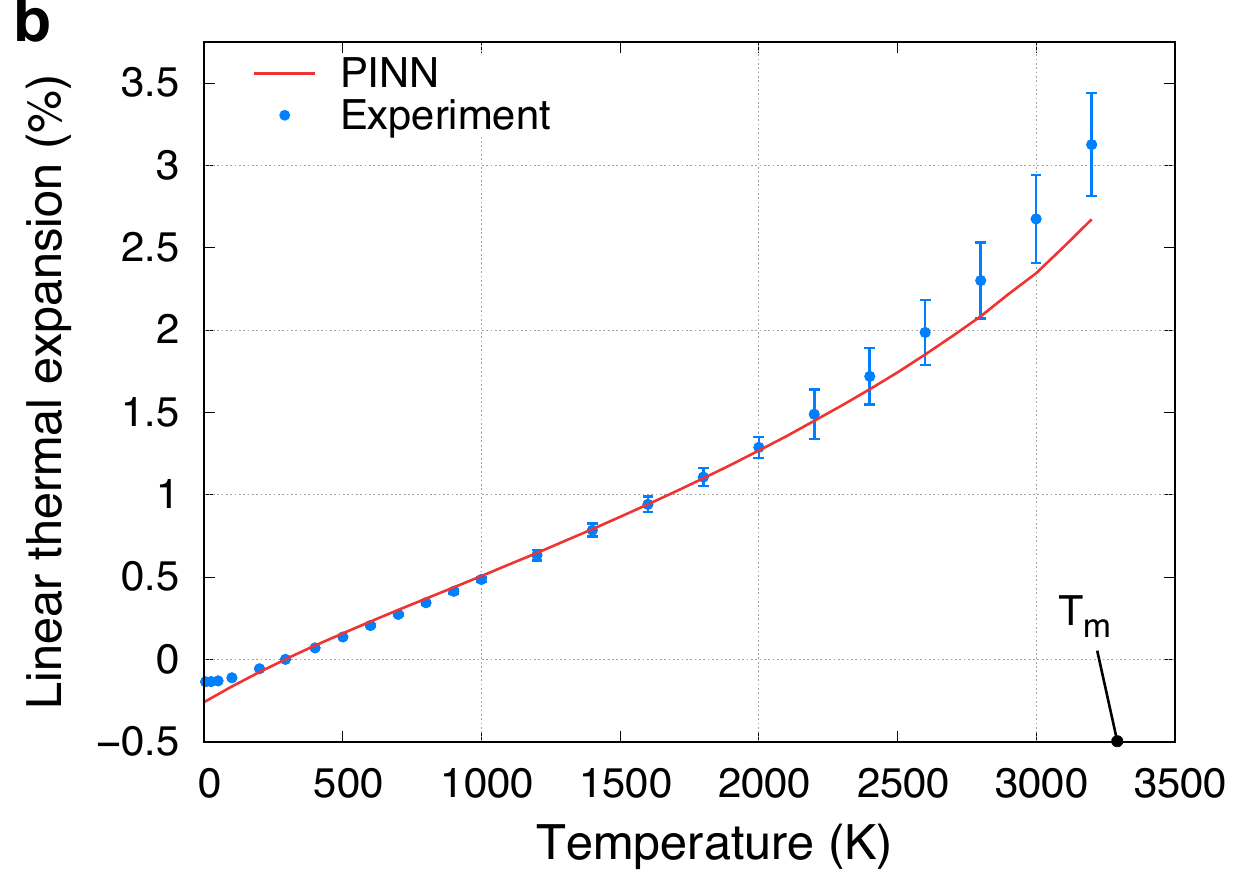}
\par\end{centering}
\caption{\label{fig:Phonon}Lattice properties of BCC Ta. \textbf{a} Phonon
dispersion relations computed with the PINN potential (red lines)
in comparison with DFT calculations (green lines) and experimental
measurements at 296 K \citep{Woods1964} (blue points). \textbf{b}
Linear thermal expansion relative to room temperature (293 K) predicted
by the PINN potential in comparison with experimental data \citep{Expansion}.
The experimental melting temperature (3293 K) $T_{\textnormal{m}}$
is indicated. The discrepancy below room temperature arises from quantum
effects that cannot be captured by a classical potential.}
\end{figure}

\begin{figure}
\noindent \begin{centering}
\includegraphics[width=0.5\columnwidth]{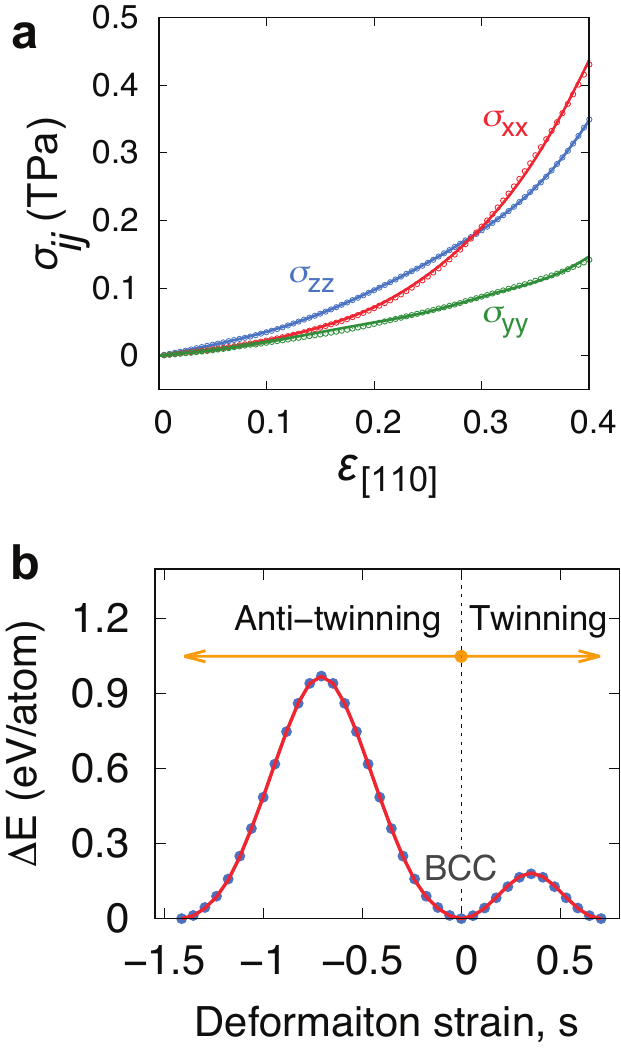}
\par\end{centering}
\caption{\label{fig:DefPath-1}BCC Ta under strong deformations. The lines
represent PINN calculations, the points DFT calculations. \textbf{a}
Stress-strain relation under strong uniaxial compression in the {[}110{]}
direction. The magnitudes of the stress components $\sigma_{xx}$,
$\sigma_{yy}$ and $\sigma_{zz}$ are plotted as a function of the
strain $\varepsilon_{xx}$. Axes orientations: $x$: $[110]$, $y$:
$[\overline{1}10]$, $z$: $[001]$. \textbf{b} Strong shear deformation.
The energy $\Delta E$ (relative to perfect BCC structure) is plotted
as a function of the strain parameter $s$ transforming the BCC structure
back to itself along the twinning and anti-twinning paths. The shear
is parallel to the $(112)$ plane in the $[11\overline{1}]$ direction.}
\end{figure}

\textbf{Properties of BCC Ta. }The Ta properties predicted by the
PINN potential were computed mainly with the ParaGrandMC (PGMC) code
\citep{ParaGrandMC,Purja-Pun:2015aa,Yamakov:2015aa} and compared
with DFT calculations performed in this work. For consistency, the
property calculations used the same density functional and other DFT
parameters as for the reference database generation. Table \ref{tab:Properties}
summarizes the basic properties of BCC Ta. The potential predicts
the equilibrium lattice constant and elastic moduli in good agreement
with DFT calculations. The phonon dispersion relations also compare
well with DFT calculations and the experimental data (Figure \ref{fig:Phonon}).
Fig.~\ref{fig:DefPath-1} demonstrates the performance of the potential
under strong lattice deformations. The PINN predictions closely match
the DFT stress-strain relations up to 40\% of the uniaxial compression
in the {[}110{]} direction (Fig.~\ref{fig:DefPath-1}a). Equally
good agreement is found for compression along the {[}100{]} and {[}111{]}
axes (Supplementary Fig.~2). This agreement makes the potential suitable
for atomistic simulations of shock deformation of Ta. As another test,
the BCC lattice was sheared in the $[11\overline{1}]$ direction parallel
to the $(112)$ plane, transforming the BCC lattice to itself along
the twinning or anti-twinning deformation paths. The energy variation
along both paths accurately follows the DFT calculations (Fig.~\ref{fig:DefPath-1}b),
which demonstrates the ability of the potential to model deformation
twinning in Ta. 

\begin{figure}
\noindent \begin{centering}
\includegraphics{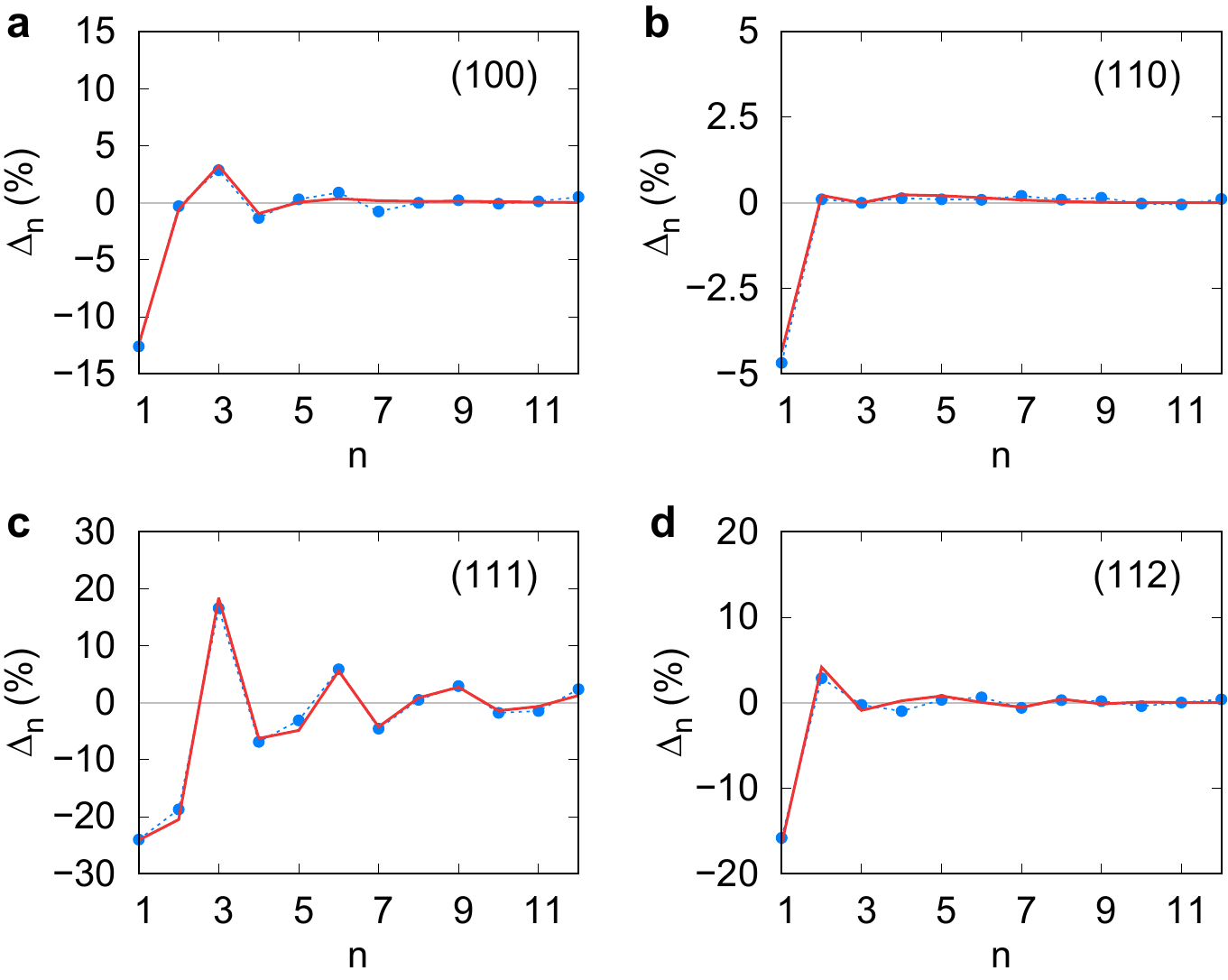}
\par\end{centering}
\caption{\label{fig:Surface}Ta surface relaxations. Atomic plane displacements
near low-index surfaces in BCC Ta predicted by the PINN potential
(red lines) in comparison with DFT calculations (blue points connected
by dash lines). $\Delta_{n}$ is the percentage change of the $n$-th
interplanar spacing relative to the ideal spacing.\textbf{ a} $\left(100\right)$,
\textbf{b} $\left(110\right)$, \textbf{c} $\left(111\right)$,\textbf{
}and \textbf{d} $\left(112\right)$\textbf{ }surface planes.}
\end{figure}

The potential accurately reproduces the DFT vacancy formation and
migration energies (Table \ref{tab:Properties}). We tested five different
configurations of self-interstitial defects and found excellent agreement
between the PINN and DFT energies in all cases. Both PINN and DFT
predict the {[}111{]} dumbbell to be the lowest energy state. Due
to its ability to describe the point-defect properties on the DFT
level of accuracy, the potential can be safely used to model diffusion-controlled
processes and radiation damage in Ta. The surface energies predicted
by the potential are close to the DFT values (Table \ref{tab:Properties}).
Furthermore, Fig.~\ref{fig:Surface} and Supplementary Fig.~3 show
that the potential faithfully reproduces the surface relaxations for
all four crystallographic orientations tested here. Surface properties
of Ta are important for catalytic applications and the biocompatibility
of porous structures \citep{Balla:2010aa}.

\begin{figure}
\noindent \begin{centering}
\includegraphics[width=0.9\columnwidth]{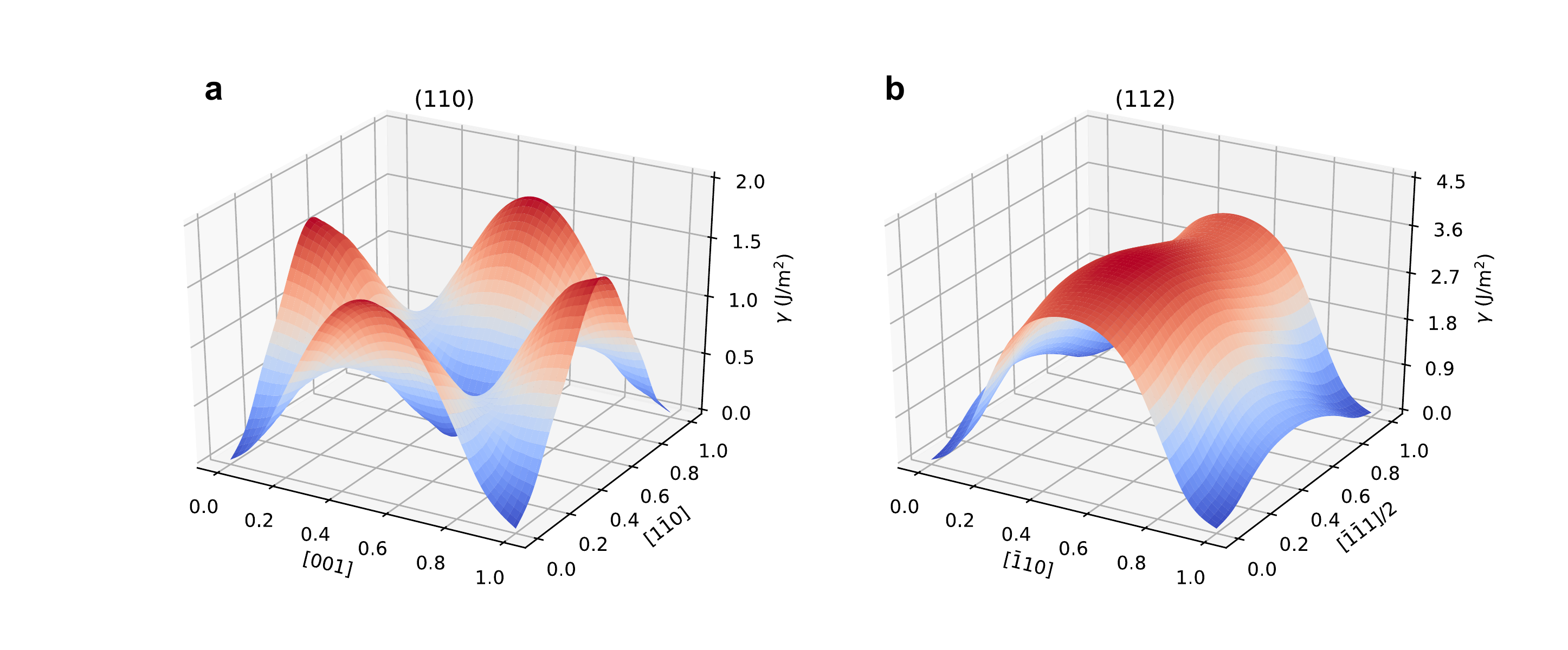}
\par\end{centering}
\noindent \begin{centering}
\includegraphics[width=0.75\columnwidth]{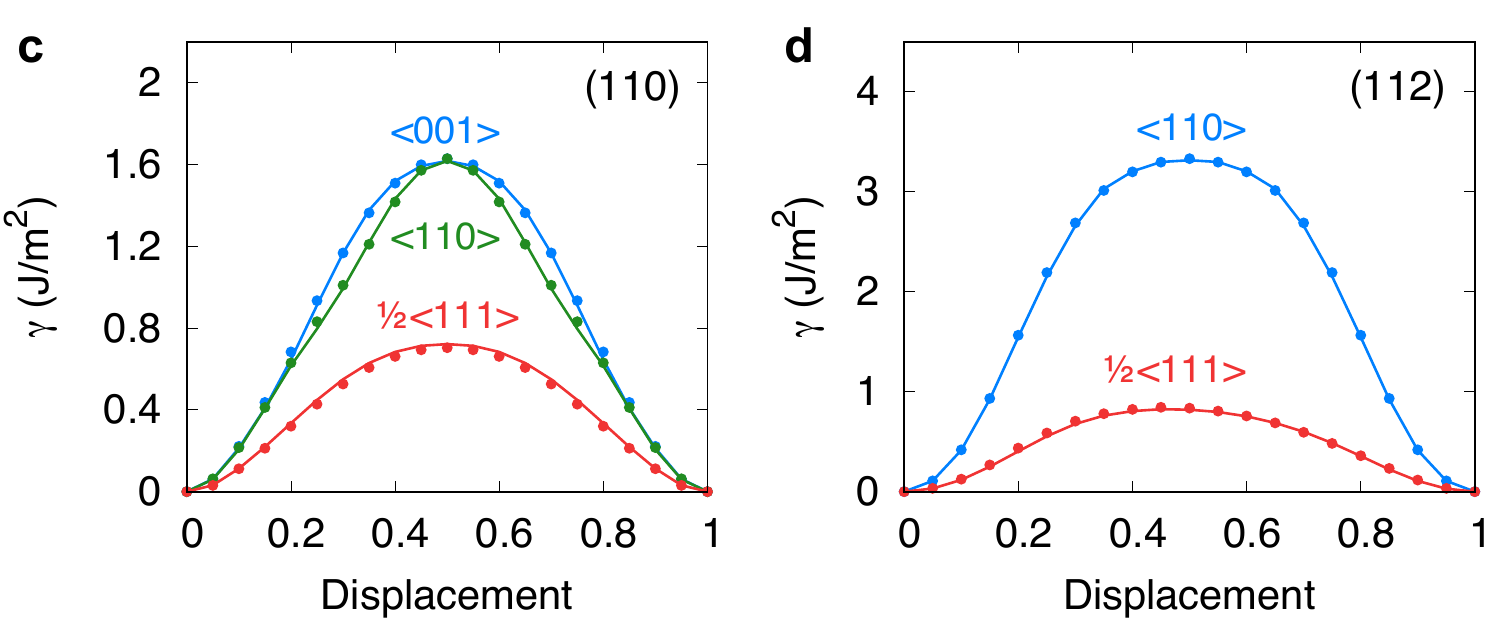}
\par\end{centering}
\caption{\label{fig:gamma}$\gamma$-surfaces in BCC Ta. \textbf{a},\textbf{b}
DFT $\gamma$-surfaces on the \textbf{a} (110) and \textbf{b} (112)
planes. \textbf{c},\textbf{d} Cross-sections of the $\gamma$-surfaces
on the \textbf{c} (110) and \textbf{d} (112) planes predicted by the
PINN potential (lines) in comparison with DFT calculations (points).
The displacements are normalized by the energy period, and their directions
are indicated next to the curves. }
\end{figure}

\begin{figure}
\noindent \begin{centering}
\includegraphics[width=0.6\columnwidth]{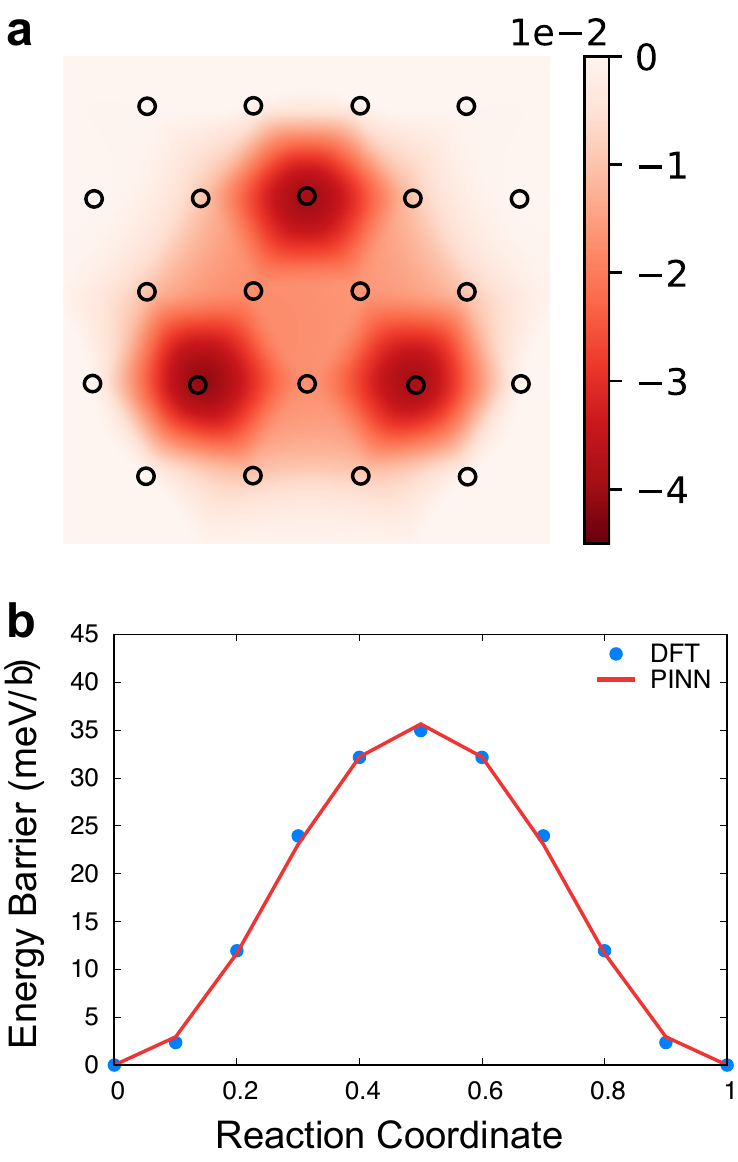}
\par\end{centering}
\caption{\label{fig:Nye} $\frac{1}{2}\left\langle 111\right\rangle $ screw
dislocation in Ta. \textbf{a} Nye tensor \citep{Hartley05} plot (the
screw component $\alpha_{zz}$) of the dislocation core structure
predicted by the PINN potential. The coordinate system is $x$: $\left[\bar{1}2\bar{1}\right]$,
$y$: $\left[\bar{1}01\right]$, $z$: $[111]$ (normal to the page).
The circles represent relaxed atomic positions in three consecutive
$\left(111\right)$ planes. \textbf{b} Peierls barrier of the dislocation
predicted by the PINN potential (lines) in comparison with DFT calculations
(points). }
\end{figure}

Mechanical behavior of Ta is controlled by the $\frac{1}{2}[111]$
screw dislocations, which have low mobility caused by the non-planar
core structure \citep{Vitek04a,CaiBCLY04,Lin2014}. In turn, the core
structure depends on the shape of the $\gamma$-surface, representing
the energy of a generalized stacking fault parallel to a given crystal
plane as a function of the displacement vector. For the screw dislocations
in BCC metals, the relevant fault planes are (110) and (211). DFT
$\gamma$-surfaces for these two planes were included in the reference
database during the potential development (Fig.~\ref{fig:gamma}a,b).
The potential accurately reproduces both $\gamma$-surfaces, as illustrated
by select cross-sections shown in Fig.~\ref{fig:gamma}c,d. The cross-sections
along the {[}111{]} direction are especially important as they determine
the lowest-energy core structure. The computed cross-sections predict
a non-generate core. To test this prediction, we directly computed
the relaxed dislocation core structure using both the potential and
the DFT (see the Methods section). The Nye tensor plot \citep{Hartley05}
in Fig.~\ref{fig:Nye}a confirms that the core computed with the
PINN potential indeed has a non-generate structure. The Nye tensor
plot obtained by DFT calculations is not shown as it looks virtually
identical to Fig.~\ref{fig:Nye}a. Non-degenerate core structures
were also found in previous DFT calculations for BCC transition metals
\citep{Weinberger:2013aa}. By contrast, most of the traditional potentials
incorrectly predict a degenerate core with a different symmetry. Many
of them also predict a spurious local minimum on the $\gamma$-surface
\citep{Moller:2018aa}.

The dislocation glide mobility depends on the Peierls barrier, defined
as the energy barrier that a straight dislocation must overcome to
move between two adjacent equilibrium positions. We computed the Peierls
barrier of the dislocation using the same supercell setup as in the
dislocation core calculations. The minimum energy path between the
fully-relaxed initial and final dislocation positions was obtained
by the nudged elastic band (NEB) method \citep{Jonsson98,HenkelmanJ00}.
The PINN and DFT calculations give nearly identical results for the
energy along the transition path, showing a single maximum in the
middle (Fig.~\ref{fig:Nye}b). The height of the barrier, normalized
by the Burgers vector's magnitude, is in close agreement with previous
DFT calculations \citep{Weinberger:2013aa}. 

\medskip{}

\begin{figure}
\noindent \begin{centering}
\includegraphics[width=0.6\columnwidth]{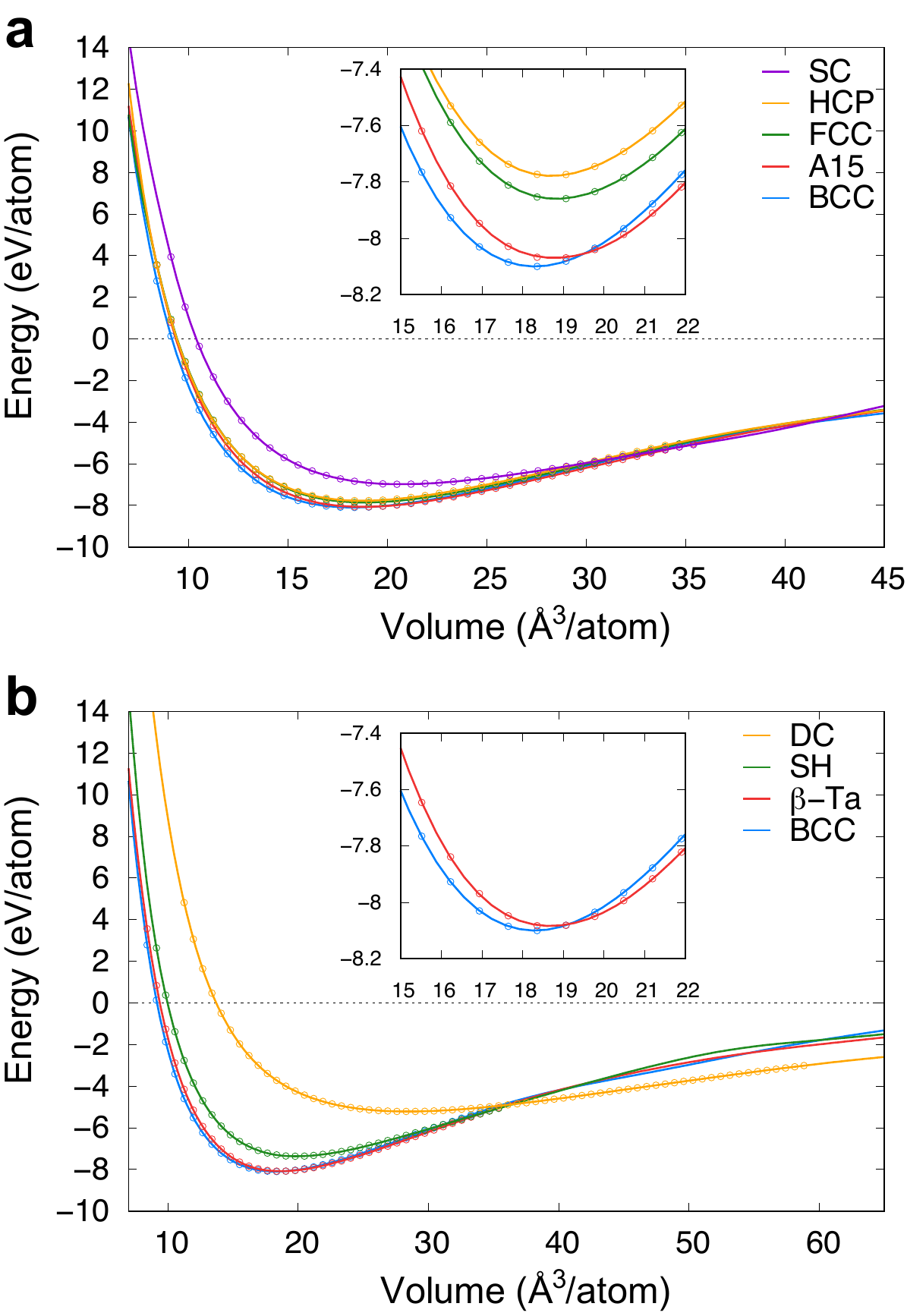}
\par\end{centering}
\caption{\label{fig:EOS}Equations of state of several crystal structures of
Ta predicted by the PINN potential (lines) in comparison with DFT
calculations (points). The insets are zooms into competing structures
near the equilibrium volume. \textbf{a} Simple cubic (SC), hexagonal
close-packed (HCP), face-centered cubic (FCC), A15 (Cr$_{3}$Si prototype),
and body-centered cubic (BCC) structures. \textbf{b} Diamond cubic
(DC), simple hexagonal (SH), $\beta$-Ta, and BCC structures.}
\end{figure}

\textbf{Alternate crystal structures of Ta.} In addition to the ground-state
BCC structure, the reference database contained the equations of state
(energy versus atomic volume) for seven more crystal structures. The
PINN potential accurately reproduces the DFT equations of state for
all seven structures, including the low-energy A15 and $\beta$-Ta
structures that compete with the BCC structure for the ground state
(Fig.~\ref{fig:EOS}). Furthermore, the potential continues to predict
physically reasonable behavior well outside the volume range covered
by the DFT points. The energy continues to rise monotonically under
strong compression and smoothly approaches zero at the cutoff distance
between atoms (Supplementary Fig.~4). This behavior illustrates the
physics-based transferability of the PINN model guided by the BOP
potential. 

Table \ref{tab:EneUnkownStruc} shows the results of additional tests.
The potential predicts the ground state energies of a dimer, a trimer,
and a tetrahedron in close agreement with DFT calculations, even though
these clusters were represented in the reference database by a small
number of supercells. As another transferability test, we performed
DFT and PINN calculations of equilibrium energies for four relatively
open crystal structures, including the A7 structure typical of nonmetals.
Although the DFT energies of these structures were not included in
the reference database, they are predicted by the potential reasonably
well (Table \ref{tab:EneUnkownStruc}). 

\begin{figure}
\noindent \begin{centering}
\includegraphics[width=0.9\columnwidth]{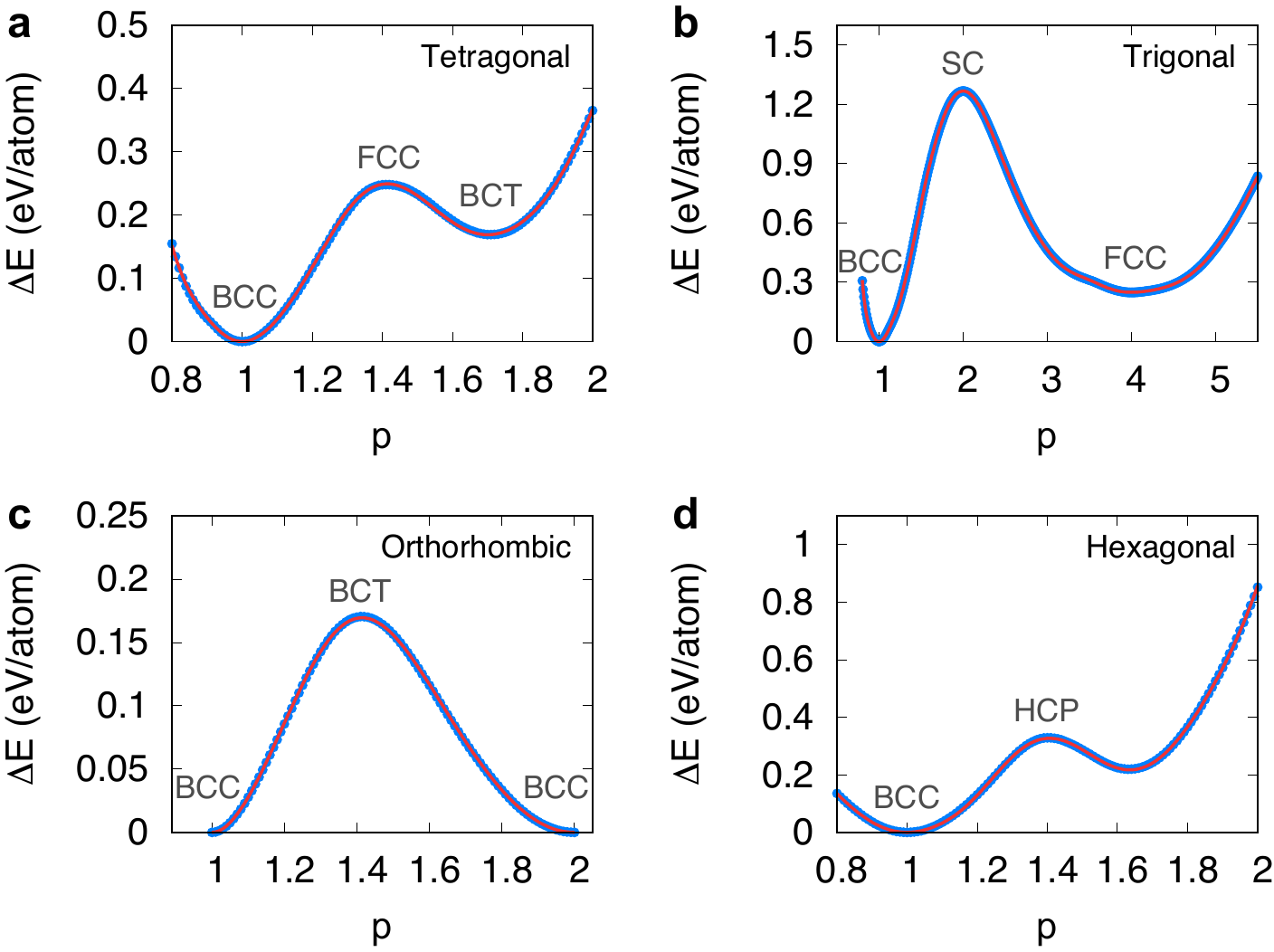}
\par\end{centering}
\caption{\label{fig:DefPath}Deformation paths between different crystal structures
of Ta. The energy $\Delta E$ (relative to perfect BCC structure)
is plotted as a function of the deformation parameter $p$ defined
in Ref.~\citep{Lin2014}. The lines represent PINN calculations,
the points DFT calculations. \textbf{a} Tetragonal (Bain) path BCC
$\rightarrow$ FCC $\rightarrow$ BCT. \textbf{b} Trigonal path BCC
$\rightarrow$ SC $\rightarrow$ FCC. \textbf{c} Orthorhombic path
BCC $\rightarrow$ BCT. \textbf{d} Hexagonal path BCC $\rightarrow$
HCP. The structures encountered along the paths are indicated.}
\end{figure}

To sample atomic configurations away from the stable and metastable
crystal structures, we included four volume-conserving deformation
paths connecting such structures. The tetragonal, trigonal, orthorhombic,
and hexagonal paths continuously transform the BCC structure to FCC,
HCP, SC, or body-centered tetragonal (BCT) structures (Fig.~\ref{fig:DefPath}).
Each path is characterized by a single deformation parameter $p$
defined in Ref.~\citep{Lin2014}. The hexagonal path combines a homogeneous
deformation with a simultaneous shuffling of alternate close-packed
atomic planes, while the remaining paths are fully homogeneous. Fig.~\ref{fig:DefPath}
shows that the PINN potential accurately reproduces the energy along
all four paths. 

\medskip{}

\textbf{The liquid phase and solid-liquid coexistence.} We computed
the liquid Ta structure by NVT MD simulations using the PINN potential
and DFT at three temperatures: below, above, and near the melting
point (see the Methods section for computational details). The potential
perfectly reproduces the radial distribution function and the bond-angle
distribution, as demonstrated in Fig.~\ref{fig:RDF} for the temperature
of 3500 K and in Supplementary Figs.~5 and 6 for two other temperatures.

The melting point calculation used the solid-liquid phase coexistence
method \citep{Morris02,Pun09b,Purja-Pun:2015aa,Howells:2018aa}. A
simulation block containing the solid and liquid phases separated
by a planar interface was coupled to a thermostat. In the canonical
ensemble, one of the phases always grows at the expense of the other,
depending on the temperature. MD simulations were performed at several
temperatures around the expected melting point, and the energy increase
or decrease rate was measured at each temperature (see the Methods
section). Interpolation to the zero rate gave us the melting temperature
$T_{m}$ at which the phases coexist in equilibrium (Supplementary
Fig.~7). The PINN potential predicts the melting temperature of 3000$\pm$6
K, which is reasonable but below the experimental value (3293 K).
Perfect agreement with experiment was not expected since the potential
was trained on DFT data without any experimental input. The only DFT
calculation that we could find in the literature gave $T_{m}=3085\pm130$
K \citep{Taioli:2007aa}, which matches our result within the statistical
uncertainty.

The liquid surface tension of Ta was computed by the capillary fluctuation
method (see Methods). The number obtained, $\gamma=1.69$ J\,m$^{-2}$,
is comparable to the experimental values of 2.07 J\,m$^{-2}$ (from
equilibrium shape of molten samples in microgravity) \citep{Miiller:1993aa}
and 2.15 J\,m$^{-2}$ from electrostatic levitation \citep{Paradis:2005aa}.
The experimental measurements were performed on relatively large (a
few mm) droplets and their accuracy was limited due to many factors,
such as temperature control, surface contamination, and evaporation.
The accurate value of $\gamma$ reported here is one of the missing
material parameters in the models of 3D printing (additive manufacturing)
of Ta and Ta-based alloys \citep{Sungail:2020aa}.

\subsection*{Discussion}

After the initial publication \citep{PINN-1}, the PINN model has
been modified by reducing the role of the NN regression to only predicting
local corrections to a global BOP potential. This step has helped
further improve the transferability of PINN potentials to atomic configurations
lying outside the reference database. As with any model, a PINN potential
eventually fails when taken too far away from the familiar territory.
However, the physics-guided extrapolation significantly expands the
potential's applicability domain compared with purely mathematical
ML models. The proposed integration of ML with a physics-based interatomic
potential preserves the DFT level of the training accuracy without
increasing the number of fitting parameters. As recently demonstrated
\citep{Pun:2020aa}, the computational overhead due to incorporating
the BOP is close to 25\%, which is marginal given that the potential
is orders of magnitude faster than straight DFT calculations.

The improved transferability of the PINN model enables the development
of general-purpose ML potentials intended for almost any type of MD
or MC simulations, similar to the off-the-shelve traditional potentials.
This capability has already been demonstrated by developing a general-purpose
PINN potential for FCC Al \citep{Pun:2020aa}. The present work has
made the next step by extending the PINN model to BCC metals. While
FCC metals can be described by EAM, MEAM, or ADP potentials reasonably
well, BCC transition metals have been notoriously challenging. For
example, most traditional potentials fail to predict the correct dislocation
core structure and the Peierls barrier in BCC metals even qualitatively.
The PINN Ta potential developed here reproduces both in excellent
agreement with DFT calculations. The potential has been trained on
a highly diverse DFT database and reproduces a broad spectrum of physical
properties of Ta with DFT accuracy. Extrapolation of the potential
to unknown atomic configurations has also been demonstrated. We foresee
that this potential will be widely used in atomistic simulations of
Ta, especially simulations of mechanical behavior under normal conditions
and shock deformation. Using the same approach, PINN potentials for
other BCC metals can be developed in the future.

The PINN method undergoes rapid development. PINN potentials for other
metallic and nonmetallic elements are now being constructed. A multi-component
version of PINN has been developed and will be reported in forthcoming
publications, together with several PINN potentials for binary systems.
The PGMC MD/MC code \citep{ParaGrandMC,Purja-Pun:2015aa,Yamakov:2015aa}
now works with multi-component PINN potentials. PINN potentials will
soon be incorporated in the Large-scale Atomic/Molecular Massively
Parallel Simulator (LAMMPS) \citep{Plimpton95}. 

\section*{Methods}

\textbf{DFT calculations.} We used the generalized gradient approximation
with the Perdew-Burke-Ernzerhof (PBE) density functional \citep{PerdewCVJPSF92,PerdewBE96}
implemented in VASP \citep{Kresse1996,Kresse1999}. The calculations
were performed with the kinetic energy cutoff of 410 eV and a Methfessel-Paxton
smearing of order 1 with the smearing width of 0.1 eV. Monkhorts-Pack
$k$-point meshes were used to sample the Brillouin zone, with the
number of grid points chosen to ensure the energy convergence to around
1 meV/atom. Before the training, all supercell energies (per atom)
were shifted by a constant such that the DFT energy of the equilibrium
BCC structure is equal to the negative experimental cohesive energy
of BCC Ta.

The phonon calculations utilized the \textsf{phonopy} package \citep{phonopy}
with a $6\times6\times6$ primitive BCC supercell (216 atoms) and
a $6\times6\times6$ Monkhorst-Pack $k$-point grid. It was verified
that, with this grid, the supercell energy was converged to 1 meV/atom,
and the phonon dispersion curves did not show noticeable changes with
the number of $k$-points. For point defects, we used a cubic supercell
of 128$\pm$1 atoms with full relaxation of both atomic positions
and the volume. The vacancy migration energy was calculated by the
NEB method \citep{Jonsson98,HenkelmanJ00} with fully relaxed supercells
as the initial and final vacancy configurations. For computing the
surface energies and surface relaxations, we used stacks of 24 atomic
layers with a chosen crystallographic orientation. The $\gamma$-surface
calculations employed similar 24-layer stacks. A generalized stacking
fault was formed in the middle of the supercell by displacing the
upper half of the layers relative to the lower half by small increments.
After each increment, the energy was minimized by only allowing relaxations
perpendicular to the fault plane. 

To find the dislocation core structure, we created a dipole configuration
consisting of 231 atoms with periodic boundary conditions in all three
dimensions as described in Ref.~\citep{Li:2004aa}. By contrast to
a single dislocation, this dipole configuration avoids the possible
incompatibility with periodic boundary conditions. The two dislocations
were first introduced into the supercell by displacing the atoms according
to the anisotropic elasticity solution for the strain field. The atomic
positions were then relaxed before examining the core structure.

The liquid structure calculations utilized a 250-atom cubic supercell
with the experimental liquid density \citep{Vinet:1993aa} near the
melting temperature. The initial BCC structure was melted and equilibrated
at the temperature of 5,000 K before quenching it to the target temperature
with the supercell box fixed. After equilibration at the target temperature,
an NVT MD run was performed in which 100 snapshots were saved at 60
fs intervals. The saved configurations were used to compute the average
structural properties of the liquid following the procedure outlined
in Ref.~\citep{Vinet:1993aa}. For the bond-angle distribution, we
only considered atoms within a sphere corresponding to the first minimum
of the radial distribution function.

\medskip{}

\textbf{The BOP potential.} The current version of the BOP potential
underlying the PINN model was described elsewhere \citep{Pun:2020aa}
and is only briefly reviewed here for completeness. The energy assigned
to atom $i$ is given by
\begin{equation}
E_{i}=\dfrac{1}{2}\sum_{j\ne i}\left[e^{A_{i}-\alpha_{i}r_{ij}}-S_{ij}b_{ij}e^{B_{i}-\beta_{i}r_{ij}}\right]f_{c}(r_{ij},d,r_{c})+E_{i}^{(p)},\label{eq:BOP1}
\end{equation}
where the summation is over neighbors $j$ separated from the atom
$i$ by a distance $r_{ij}$. The interactions are truncated at the
cutoff distance $r_{c}$ using the cutoff function
\begin{equation}
f_{c}(r,r_{c},d)=\begin{cases}
\dfrac{(r-r_{c})^{4}}{d^{4}+(r-r_{c})^{4}}\enskip & r\leq r_{c}\\
0,\enskip & r\geq r_{c},
\end{cases}\label{eq:BOP2}
\end{equation}
where the parameter $d$ controls the truncation smoothness. The first
term in the square brackets in Eq.(\ref{eq:BOP1}) describes the repulsion
between atoms at short separations. The second term describes chemical
bonding and captures the bond-order effect through the coefficient
\begin{equation}
b_{ij}=(1+z_{ij})^{-1/2},\label{eq:BOP3}
\end{equation}
where $z_{ij}$ represents the number of bonds $ik$ formed by the
atom $i$. The bonds are counted with weights depending on the bond
angle $\theta_{ijk}$:
\begin{equation}
z_{ij}=\sum_{k\ne i,j}a_{i}S_{ik}\left(\cos\theta_{ijk}-h_{i}\right)^{2}f_{c}(r_{ik},d,r_{c}).\label{eq:BOP4}
\end{equation}
The angular dependence account for the directional character of the
bonds. All chemical bonds are screened by the screening factor $S_{ij}$
defined by
\begin{equation}
S_{ij}=\prod_{k\ne i,j}S_{ijk},\label{eq:BOP5}
\end{equation}
where the partial screening factors $S_{ijk}$ represent the contributions
of individual atoms $k$ to the screening of the bond $i$-$j$:
\begin{equation}
S_{ijk}=1-f_{c}(r_{ik}+r_{jk}-r_{ij},d,r_{c})e^{-\lambda_{i}(r_{ik}+r_{jk}-r_{ij})},\label{eq:BOP6}
\end{equation}
where $\lambda_{i}$ is the inverse of the screening length. The closer
the atom $k$ is to the bond $i$-$j$, the smaller $S_{ijk}$, and
thus the larger is its contribution to the screening. Finally, the
on-site energy
\begin{equation}
E_{i}^{(p)}=-\sigma_{i}\left({\displaystyle \sum_{j\neq i}S_{ij}b_{ij}}f_{c}(r_{ij})\right)^{1/2}\label{eq:BOP7}
\end{equation}
represents the embedding energy in metals and the promotion energy
for covalent bonding.

The potential functions depend on 10 parameters, 8 of which ($A$,
$B$, $\alpha$, $\beta$, $a$, $h$, $\lambda$ and $\sigma$) are
locally adjusted by the NN while $d$ and $r_{c}$ are treated as
global.

\medskip{}

\textbf{The training procedure.} The NN weights and biases were initialized
by random numbers in the interval {[}-0.3,0.3{]}. The goal of the
training was to minimize the loss function
\begin{eqnarray}
\mathcal{E} & = & \dfrac{1}{N}\sum_{s}\left(\dfrac{E^{s}-E_{\textrm{DFT}}^{s}}{N_{s}}\right)^{2}+\tau_{1}\dfrac{1}{N_{p}}\left(\sum_{\epsilon\kappa}\left|w_{\epsilon\kappa}\right|^{2}+\sum_{\nu}\left|b_{\nu}\right|^{2}\right)\nonumber \\
 & + & \tau_{2}\dfrac{1}{N_{a}m}\sum_{s}\sum_{i_{s}=1}^{N_{s}}\sum_{n=1}^{m}\left|p_{i_{s}n}-\overline{p}_{i_{s}n}\right|^{2}\label{eq:NN3}
\end{eqnarray}
with respect to the weights $w_{\epsilon\kappa}$ and biases $b_{\nu}$.
In Eq.(\ref{eq:NN3}), $E^{s}$ is the total energy of supercell $s$
predicted by the potential, $E_{\textrm{DFT}}^{s}$ is the respective
DFT energy, $N_{s}$ is the number of atoms in the supercell $s$,
$N$ is the number of supercells in the reference database, $N_{a}$
is the total number of atoms in all supercells, $N_{p}$ is the total
number of NN parameters, and $\tau_{1}$ and $\tau_{2}$ are adjustable
coefficients. The second and third terms are added for regularization
purposes. The second term ensures that the network parameters remain
reasonably small for smooth interpolation. The third term controls
the variations of the BOP parameters relative to their values $\overline{p}_{i_{s}n}$
averaged over the training database. The values $\tau_{1}$ = $10^{-10}$
and $\tau_{2}$ = $10^{-10}$ were chosen from previous experience
\citep{Pun:2020aa}. The L-BFGS algorithm implementing the minimization
requires the knowledge of partial derivatives of $\mathcal{E}$ with
respect to the NN parameters, which were derived analytically and
implemented in the training code. Since the loss function has many
local minima, the training had to be repeated multiple times, starting
from different initial conditions. Due to the large size of the optimization
problem, the training process heavily relies on massive parallel computations.

\medskip{}

\textbf{PINN calculations of Ta properties.} The PINN calculations
utilized the same atomic configurations as in the DFT calculations,
except for larger system sizes in some cases. The phonon calculations
employed the same \textsf{phonopy} package \citep{phonopy}. The thermal
expansion was computed by NPT MD simulations on a 4,394-atom periodic
block. This simulation block was also used for point-defect and liquid-structure
calculations. A stack of 48 atomic layers was created for the surface
and $\gamma$-surface calculations. The dislocation core analysis
and the Peierls barriers calculation were based on the same supercells
as in the DFT calculations.

The melting point calculation followed the methodology of \citep{Pun:2020aa}.
An orthorhombic periodic simulation block had the dimensions of $39\times42\times202$
\AA{} (16,320 atoms) and contained approximately equal amounts of
the solid and liquid phases. The solid-liquid interface had the (111)
crystallographic orientation and was normal to the long direction
of the block. The solid phase was scaled according to the thermal
expansion factor at the chosen temperature to eliminate internal stresses.
The MD simulations were performed in the canonical ensemble, in which
the interface cross-section remained fixed while the long dimension
of the simulation block was allowed to vary to ensure zero normal
stress. The solid crystallized if the temperature was below the melting
temperature $T_{m}$ and melted if it was above $T_{m}$. The phase
change was accompanied by a decrease in the energy in the first case
and an increase in the second. The steady-state energy rate was recorded
as a function of the simulation temperature and interpolated to zero
to obtain $T_{m}$. The liquid surface tension of Ta was computed
by the capillary fluctuation method closely following the methodology
of the previous calculations for Al \citep{Pun:2020aa}. The liquid
simulation block had a ribbon-type geometry with the dimensions of
$613\times34\times212$ \AA{} (216,000 atoms) as shown in Supplementary
Fig.~8. After equilibration at the melting temperature, 241 MD snapshots
were saved at 1 ps time intervals. The capillary wave amplitudes $A(k)$
were obtained by a discrete Fourier transformation of the liquid surface
shape. The surface tension $\gamma$ was obtained from the linear
fit to the plot of the inverse of the ensemble-averaged spectral power
$\left\langle \left|A(k)\right|^{2}\right\rangle $ versus the wave-number
squared $k^{2}$ in the long wavelength limit (Supplementary Fig.~9).

\bigskip{}

\noindent \textbf{Data availability}

\noindent All data that support the findings of this study are available
in the Supplementary Information file and from the corresponding author
upon reasonable request.


\noindent \bigskip{}
\bigskip{}

\noindent \textbf{Acknowledgements}

\noindent We are grateful to Dr.~Vesselin Yamakov for developing
the PGMC software used for most of the PINN-based simulations reported
in this work. This research was supported by the Office of Naval Research
under Award No.~N00014-18-1-2612. The computations were supported
in part by a grant of computer time from the DoD High Performance
Computing Modernization Program at ARL DSRC, ERDC DSRC and Navy DSRC.
Additional computational resources were provided by the NASA High-End
Computing (HEC) Program through the NASA Advanced Supercomputing (NAS)
Division at Ames Research Center.

\bigskip{}

\noindent \textbf{Author contributions}

\noindent Y.-S.\ L.\ performed all DFT calculations for the Ta database
and Ta properties, developed the computer software for the PINN potential
training, created the PINN Ta potential, and performed all PINN-based
simulations (except for the liquid surface tension) under Y.\ M.'s
direction and supervision. G.~P.~P.P.~shared his computer scripts
and experience with PINN training and simulations, and computed the
liquid surface tension reported in this work. Y.\ M.\ wrote the
initial draft of the manuscript, which was then edited and approved
in its final form by all authors.

\bigskip{}

\noindent \textbf{Competing interests}

\noindent The authors declare no competing interests.

\clearpage{}

\begin{table}
\caption{\label{tab:Properties}Properties of Ta predicted by the PINN potential,
DFT calculations, and measured by experiment. }

\centering{}%
\begin{tabular}{lcccccccc}
\toprule 
 & DFT (this work) &  & PINN &  &  & DFT (others' work) &  & Experiments\tabularnewline
\midrule
\midrule 
$a_{0}$ (\AA ) & 3.3202 &  & 3.3203 &  &  & 3.321$^{a}$ &  & 3.3039$^{b}$ \tabularnewline
$E_{\mathrm{coh}}$ (eV/atom) &  &  & 8.1 &  &  &  &  & 8.1$^{c}$\tabularnewline
$B$ (GPa) & 196 &  & 198 &  &  &  &  & 194.2$^{d}$; 198$^{e}$ \tabularnewline
$C_{11}$ (GPa) & 268 &  & 269 &  &  & 247$^{a}$  &  & 266.3$^{d}$ \tabularnewline
$C_{12}$ (GPa) & 159 &  & 163 &  &  & 170$^{a}$  &  & 158.2$^{d}$ \tabularnewline
$C_{44}$ (GPa) & 73 &  & 72 &  &  & 67$^{a}$  &  & 87.4$^{d}$ \tabularnewline
$E_{v}^{f}$ (eV) & 2.866 &  & 2.772 &  &  & 2.99$^{f}$; 2.95$^{g}$; 2.91$^{i}$  &  & 2.2--3.1$^{h}$ \tabularnewline
$E_{v}^{m}$ (eV) & 0.77 &  & 0.76 &  &  & 0.83$^{f}$; 0.76$^{j}$  &  & 0.7$^{h}$ \tabularnewline
$E_{i}^{f}\left\langle 111\right\rangle $ dumbbell (eV) & 4.730 &  & 4.687 &  &  & 5.832$^{k}$  &  & \tabularnewline
$E_{i}^{f}\left\langle 110\right\rangle $ dumbbell (eV) & 5.456 &  & 5.414 &  &  & 5.836$^{k}$  &  & \tabularnewline
$E_{i}^{f}$ tetrahedral (eV) & 5.799 &  & 5.663 &  &  & 6.771$^{k}$  &  & \tabularnewline
$E_{i}^{f}\left\langle 100\right\rangle $ dumbbell (eV) & 5.892 &  & 5.820 &  &  & 7.157$^{k}$  &  & \tabularnewline
$E_{i}^{f}$ octahedral (eV) & 5.940 &  & 5.986 &  &  & 7.095$^{k}$  &  & \tabularnewline
$\gamma_{s}\left(110\right)$ $(\mathrm{J}/\mathrm{m}^{2})$ & 2.369 &  & 2.367 &  &  & 2.31$^{l}$; 2.881$^{j}$; 2.51$^{m}$  &  & 2.49$^{n*}$\tabularnewline
$\gamma_{s}\left(100\right)$ $(\mathrm{J}/\mathrm{m}^{2})$ & 2.483 &  & 2.482 &  &  & 2.27$^{l}$; 3.142$^{j}$; 2.82$^{m}$ &  & \tabularnewline
$\gamma_{s}\left(112\right)$ $(\mathrm{J}/\mathrm{m}^{2})$ & 2.675 &  & 2.671 &  &  & 2.71$^{l}$; 3.270$^{j}$  &  & \tabularnewline
$\gamma_{s}\left(111\right)$ $(\mathrm{J}/\mathrm{m}^{2})$ & 2.718 &  & 2.718 &  &  & 2.74$^{l}$; 3.312$^{j}$  &  & \tabularnewline
$\gamma_{\mathrm{us}}\left\{ 110\right\} \left\langle 001\right\rangle $
$(\mathrm{J}/\mathrm{m}^{2})$ & 1.628 &  & 1.618 &  &  & 1.951$^{a}$ &  & \tabularnewline
$\gamma_{\mathrm{us}}\left\{ 110\right\} \left\langle 110\right\rangle $
$(\mathrm{J}/\mathrm{m}^{2})$ & 1.628 &  & 1.618 &  &  &  &  & \tabularnewline
$\gamma_{\mathrm{us}}\left\{ 110\right\} \left\langle 111\right\rangle $
$(\mathrm{J}/\mathrm{m}^{2})$ & 0.704 &  & 0.722 &  &  & 0.840$^{a}$ &  & \tabularnewline
$\gamma_{\mathrm{us}}\left\{ 112\right\} \left\langle 110\right\rangle $
$(\mathrm{J}/\mathrm{m}^{2})$ & 3.330 &  & 3.315 &  &  &  &  & \tabularnewline
$\gamma_{\mathrm{us}}\left\{ 112\right\} \left\langle 111\right\rangle $
$(\mathrm{J}/\mathrm{m}^{2})$ & 0.844 &  & 0.825 &  &  & 1.000$^{a}$ &  & \tabularnewline
$T_{m}$ (K) &  &  & 3000$\pm$6 &  &  & 3085$\pm$130$^{o}$ &  & 3293$^{c}$ \tabularnewline
\midrule
\multicolumn{9}{l}{$a_{0}$ equilibrium lattice constant, $E_{\mathrm{coh}}$ equilibrium
cohesive energy, $B$ bulk modulus, }\tabularnewline
\multicolumn{9}{l}{$C_{ij}$ elastic constants, $E_{v}^{f}$ vacancy formation energy,
$E_{v}^{m}$ vacancy migration energy, }\tabularnewline
\multicolumn{9}{l}{$E_{i}^{f}$ self-interstitial formation energy, $\gamma_{s}$ surface
energy, $\gamma_{\mathrm{us}}$ unstable stacking fault energy, }\tabularnewline
$T_{m}$ melting temperature. &  &  &  &  &  &  &  & \tabularnewline
\multicolumn{9}{l}{$^{a}$Ref.~\citep{Ravelo2013}, $^{b}$Ref.~\citep{Dewaele2004},
$^{c}$Ref.~\citep{Kittel}, $^{d}$Ref.~\citep{Featherston:1963aa}, }\tabularnewline
\multicolumn{9}{l}{$^{e}$Ref.~\citep{Dewaele2004}, $^{f}$Ref.~\citep{Satta1999},
$^{g}$Ref.~\citep{Mukherjee2003}, $^{j}$Ref.~\citep{Mishin.Ta},}\tabularnewline
\multicolumn{9}{l}{$^{h}$Ref.~\citep{LandoltIII.25}, $^{i}$Ref.~\citep{Feng2011},$^{k}$Ref.~\citep{Nguyen-Manh2006},
$^{l}$Ref.~\citep{Kiejna2005},}\tabularnewline
\multicolumn{9}{l}{$^{m}$Ref.~\citep{Wu1995}, $^{n}$Ref.~\citep{Tyson:1977aa},
$^{o}$Ref.~\citep{Taioli:2007aa}, $^{*}$Average orientation }\tabularnewline
\end{tabular}
\end{table}

\begin{table}
\caption{\label{tab:EneUnkownStruc} Equilibrium energy (in eV/atom) of small
clusters and open crystal structures unknown to the PINN potential.
The values are reported relative to the equilibrium BCC structure. }
\bigskip{}

\begin{tabular}{llcccc}
\hline 
Structure &  & DFT &  &  & PINN\tabularnewline
\hline 
Dimer &  & 6.044 &  &  & 6.048\tabularnewline
Trimer (linear) &  & 5.944 &  &  & 5.947\tabularnewline
Trimer (triangle) &  & 4.754 &  &  & 4.756\tabularnewline
Tetrahedron &  & 3.667 &  &  & 3.668\tabularnewline
\hline 
$\alpha$-U (A20) &  & 0.375 &  &  & 0.381\tabularnewline
$\beta$-Sn (A5) &  & 0.768 &  &  & 0.808\tabularnewline
$\alpha$-Ga (A11) &  & 1.300 &  &  & 1.283\tabularnewline
$\alpha$-As (A7) &  & 1.782 &  &  & 1.580\tabularnewline
\hline 
\end{tabular}
\end{table}

\pagebreak\clearpage{}

\begin{figure}
\noindent \begin{centering}
\includegraphics[width=0.6\columnwidth]{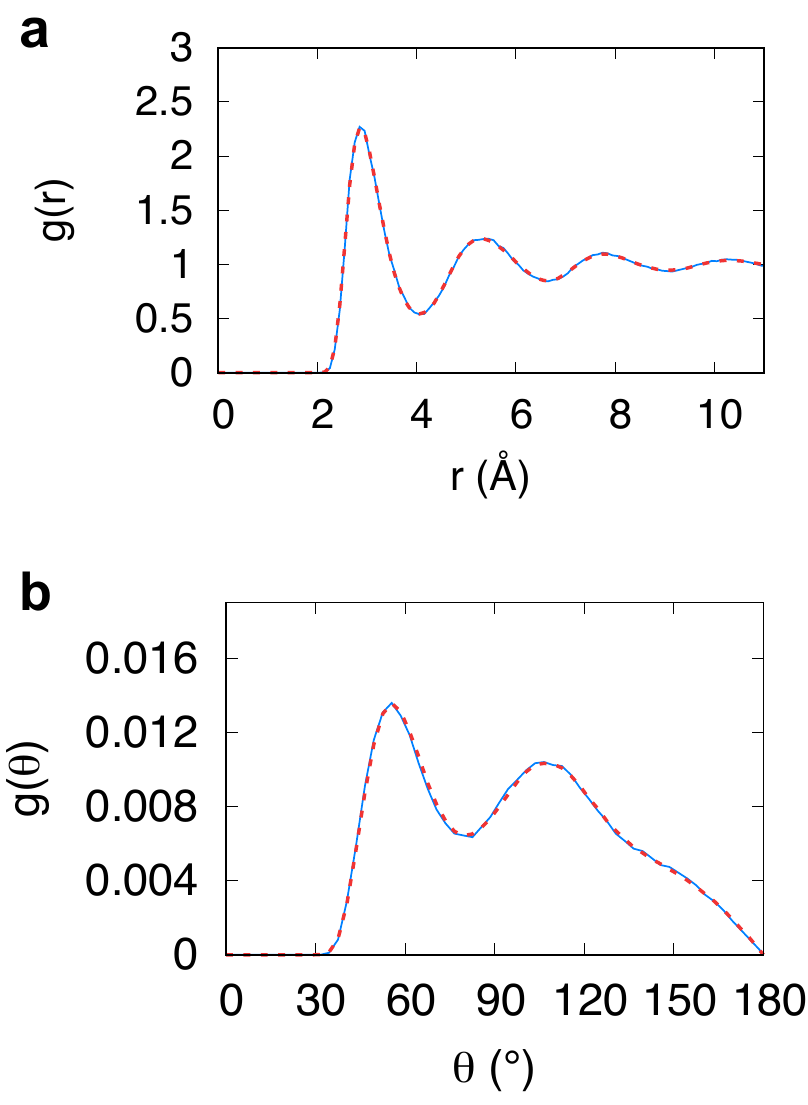}
\par\end{centering}
\caption{\textbf{\label{fig:RDF}}Ta liquid properties. Predictions of the
PINN potential (red dashed lines) are compared with DFT calculations
(blue lines) at the temperature of 3500 K. \textbf{a} Radial distribution
function. \textbf{b} Bond-angle distribution function.}
\end{figure}

\newpage\clearpage{}

\setcounter{page}{1}\date{}
\noindent \begin{center}
\textsf{\textbf{\huge{}Supplementary Information}}{\huge\par}
\par\end{center}

\begin{center}
 
\par\end{center}

\begin{center}
\bigskip{}
\par\end{center}

\noindent \begin{center}
\textbf{\LARGE{}Development of a physically-informed neural network
interatomic potential for tantalum}{\LARGE\par}
\par\end{center}

\bigskip{}
\bigskip{}

\noindent \begin{center}
\textbf{\Large{}Yi-Shen Lin, Ganga P. Purja Pun and Yuri Mishin}{\Large\par}
\par\end{center}

\setcounter{figure}{0}
\setcounter{table}{0} 

\makeatletter \renewcommand{\thefigure}{S\arabic{figure}}
\renewcommand{\thetable}{S\arabic{table}}

\begin{table}
\caption{\label{tab:DatabaseTrainValidation}Ta DFT database used for the development
of the PINN potential.}

\begin{tabular}{lllll}
\hline 
Subset & Structure & Simulation type & $N_{A}$ & $N_{tv}$\tabularnewline
\hline 
Crystals & BCC & Small homogeneous strains & 2 & 63\tabularnewline
 & BCC & Isotropic strain at 0 K & 2 & 38\tabularnewline
 & A15 & Isotropic strain at 0 K & 8 & 39\tabularnewline
 & $\beta$-Ta & Isotropic strain at 0 K & 30 & 39\tabularnewline
 & Diamond cubic & Isotropic strain at 0 K & 8 & 68\tabularnewline
 & FCC & Isotropic strain at 0 K & 4 & 38\tabularnewline
 & HCP & Isotropic strain at 0 K & 4 & 37\tabularnewline
 & SH & Isotropic strain at 0 K & 2 & 38\tabularnewline
 & SC & Isotropic strain at 0 K & 1 & 38\tabularnewline
\hline 
Small clusters & dimer & Isotropic strain at 0 K & 2 & 12\tabularnewline
 & trimer-linear & Isotropic strain at 0 K & 3 & 12\tabularnewline
 & trimer-triangle & Isotropic strain at 0 K & 3 & 12\tabularnewline
 & tetrahedron & Isotropic strain at 0 K & 4 & 12\tabularnewline
\hline 
Deformation paths & --- & twinning-antitwinning & 1 & 37\tabularnewline
 & --- & hexagonal & 4 & 121\tabularnewline
 & --- & orthorhombic & 4 & 101\tabularnewline
 & --- & tetragonal & 2 & 121\tabularnewline
 & --- & trigonal & 1 & 471\tabularnewline
\hline 
$\gamma$-surface & --- & (110) plane & 24 & 54\tabularnewline
 & --- & (112) plane & 24 & 81\tabularnewline
\hline 
Liquid & --- & NVT-MD (2600 K) & 250 & 20\tabularnewline
 & --- & NVT-MD (2900 K) & 250 & 20\tabularnewline
 & --- & NVT-MD (3500 K) & 250 & 20\tabularnewline
 & --- & NVT-MD (5000 K) & 250 & 20\tabularnewline
\hline 
BCC-1 & BCC ($a$ = 3.3202) & NVT-MD (2500 K) & 54 & 20\tabularnewline
 & BCC ($a$ = 3.3202) & NVT-MD (5000 K) & 54 & 40\tabularnewline
 & BCC ($a$ = 3.3202) & NVT-MD (10,000 K) & 54 & 40\tabularnewline
\hline 
BCC-2 & BCC ($e$ = $-$2.5\% ) & NVT-MD (2500 K) & 54 & 40\tabularnewline
 & BCC ($e$ = 2.5\% ) & NVT-MD (2500 K) & 54 & 40\tabularnewline
 & BCC ($e$ = $-$5\% ) & NVT-MD (2500 K) & 54 & 40\tabularnewline
 & BCC ($e$ = 5\% ) & NVT-MD (2500 K) & 54 & 40\tabularnewline
\hline 
\multicolumn{2}{l}{Continued in Supplementary Table \ref{tab:DatabaseTrainValidation-1}} &  &  & \tabularnewline
\end{tabular}
\end{table}

\noindent 
\begin{table}
\noindent \caption{Ta DFT database (continued from Supplementary Table \ref{tab:DatabaseTrainValidation}).\label{tab:DatabaseTrainValidation-1}}

\begin{tabular}{lllll}
\hline 
Subset & Structure & Simulation type & $N_{A}$ & $N_{tv}$\tabularnewline
\hline 
BCC-3 & BCC & Large uniaxial stain along $\left[100\right]$ & 2 & 80\tabularnewline
 & BCC & Large uniaxial stain along $\left[110\right]$ & 4 & 80\tabularnewline
 & BCC & Large uniaxial stain along $\left[111\right]$ & 6 & 79\tabularnewline
\hline 
BCC-4 & BCC ($a$ = 3.3247 ) & NVT-MD (293 K) & 128 & 180\tabularnewline
 & BCC ($a$ = 3.3292) & NVT-MD (500 K) & 128 & 210\tabularnewline
 & BCC ($a$ = 3.3247) & NVT-MD (293 K) & 54 & 140\tabularnewline
 & BCC ($a$ = 3.3292) & NVT-MD (500 K) & 54 & 140\tabularnewline
 & BCC ($a$ = 3.3408) & NVT-MD (1000 K) & 54 & 60\tabularnewline
 & BCC ($a$ = 3.3534) & NVT-MD (1500 K) & 54 & 60\tabularnewline
 & BCC ($a$ = 3.3675) & NVT-MD (2000 K) & 54 & 40\tabularnewline
 & BCC ($a$ = 3.3863) & NVT-MD (2500 K) & 54 & 40\tabularnewline
 & BCC ($a$ = 3.4137) & NVT-MD (3000 K) & 54 & 40\tabularnewline
 & BCC ($a$ = 3.4287) & NVT-MD (3200 K) & 54 & 40\tabularnewline
 & BCC ($a$ = 3.4287) & NVT-MD (3300 K) & 54 & 40\tabularnewline
\hline 
Dislocation & Dislocation & Relaxed structure and NEB calculations & 231 & 46\tabularnewline
\hline 
Spherical clusters & radius = $2^{\mathrm{nd}}$ NND & NVT-MD (2600 K) & 15 & 40\tabularnewline
 & radius = $3^{\mathrm{rd}}$ NND & NVT-MD (2500 K) & 27 & 40\tabularnewline
 & radius = $4^{\mathrm{th}}$ NND & NVT-MD (2500 K) & 51 & 40\tabularnewline
\hline 
Interfaces & GB $\Sigma3\left(111\right)$ & NVT-MD (2500 K) & 48 & 40\tabularnewline
 & GB $\Sigma3\left(112\right)$ & NVT-MD (2500 K) & 72 & 40\tabularnewline
 & GB $\Sigma5\left(210\right)$ & NVT-MD (2500 K) & 36 & 40\tabularnewline
 & GB $\Sigma5\left(310\right)$ & NVT-MD (2500 K) & 60 & 40\tabularnewline
\hline 
Point defects & Vacancy & Relaxed structure and NEB calculations & 127 & 16\tabularnewline
 & Vacancy & NVT-MD (2500 K) & 53 & 40\tabularnewline
 & Self-interstitials & Relaxed structures & 129 & 5\tabularnewline
 & Self-interstitials & NVT-MD (2500 K) & 55 & 40\tabularnewline
\hline 
Surfaces & Surfaces & Relaxed structures & 24 & 4\tabularnewline
 & Surface $\left(100\right)$ & NVT-MD (2500 K) & 63 & 40\tabularnewline
 & Surface $\left(110\right)$ & NVT-MD (2500 K) & 60 & 40\tabularnewline
 & Surface $\left(111\right)$ & NVT-MD (2500 K) & 44 & 40\tabularnewline
\hline 
\multicolumn{5}{l}{$N_{tv}$: number of configurations for training and cross-validation.}\tabularnewline
\multicolumn{5}{l}{$N_{A}$: number of atoms per supercell. $a$: cubic lattice parameter
for BCC in \AA .}\tabularnewline
\multicolumn{5}{l}{NND: nearest neighbor distance in equilibrium BCC structure.}\tabularnewline
\multicolumn{5}{l}{$e$: isotropic stain from equilibrium BCC. GB: {[}001{]} symmetric
tilt grain boundary.}\tabularnewline
\end{tabular}
\end{table}

\begin{figure}
\noindent \begin{centering}
\includegraphics[width=0.45\columnwidth]{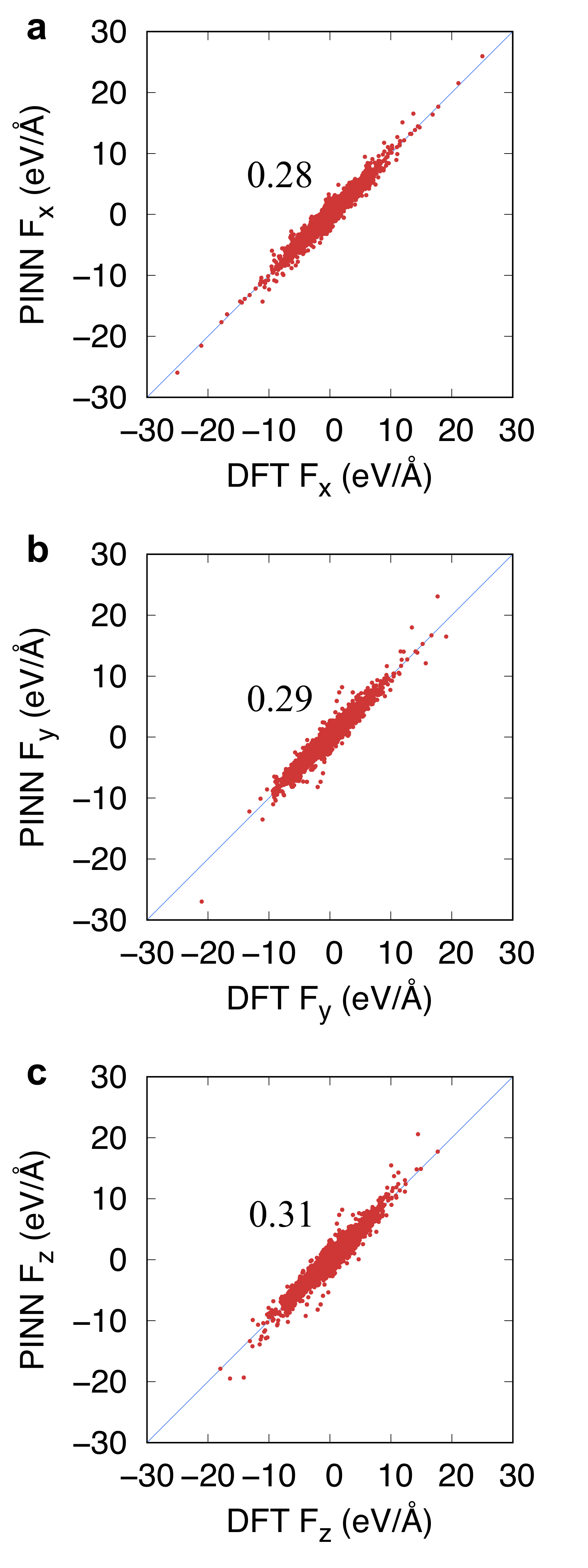}
\par\end{centering}
\caption{Testing of atomic force predictions. The components of the atomic
forces predicted by the PINN potential are compared with DFT calculations.
The straight lines represent the perfect fit. \textbf{a} $x$-components,
\textbf{b} $y$-components, and \textbf{c} $z$-components. The RMSE
in eV/\AA{} is indicated next to the points. \label{fig:forces}}
\end{figure}

\begin{figure}
\noindent \begin{centering}
\includegraphics[width=0.45\columnwidth]{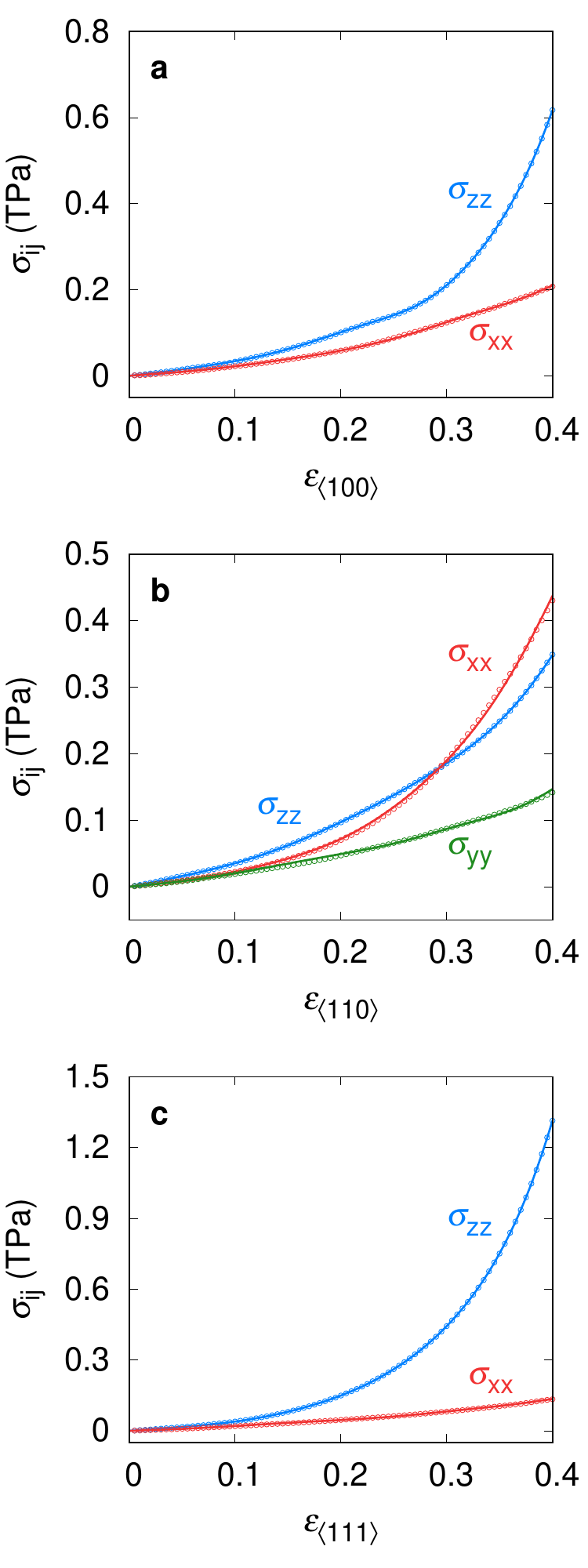}
\par\end{centering}
\caption{\label{fig:DefPath-1-1}BCC Ta under strong uniaxial compression.
The lines represent PINN calculations, the points DFT calculations.
The magnitudes of the stress components $\sigma_{xx}$, $\sigma_{yy}$
and $\sigma_{zz}$ are plotted as a function of the strain $\varepsilon_{xx}$.
Axes orientations: a $x$: $[110]$, $y$: $[\overline{1}10]$, $z$:
$[001]$. b $x$: $[100]$, $y$: $[010]$, $z$: $[001]$. c $x$:
$[111]$, $y$: $[\overline{1}10]$, $z$: $[\overline{1}\overline{1}2]$.}
\end{figure}

\begin{figure}
\noindent \begin{centering}
\includegraphics[width=0.6\columnwidth]{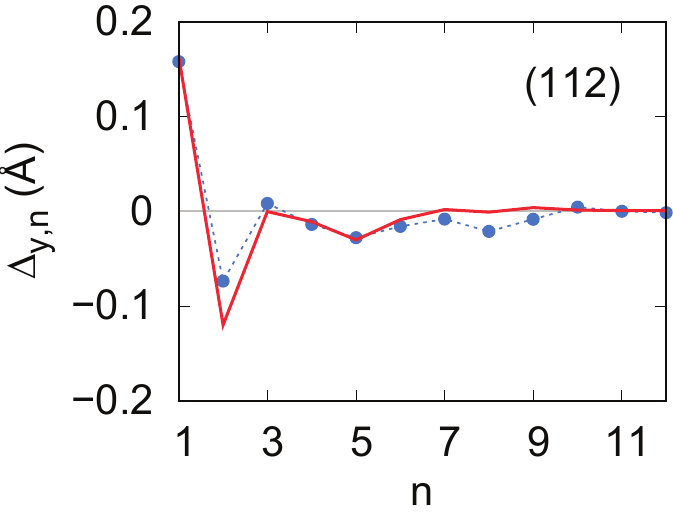}
\par\end{centering}
\caption{\label{fig:Surface-1}Ta surface relaxations. Atomic plane displacements
near the (112) surface in BCC Ta predicted by the PINN potential (red
lines) in comparison with DFT calculations (blue points connected
by dash lines). $\Delta_{y,n}$ is the displacement of $n$-th atomic
layer in the $\left[\bar{1}\bar{1}1\right]$ direction due to relaxation
from its ideal position.}
\end{figure}

\begin{figure}
\noindent \begin{centering}
\includegraphics{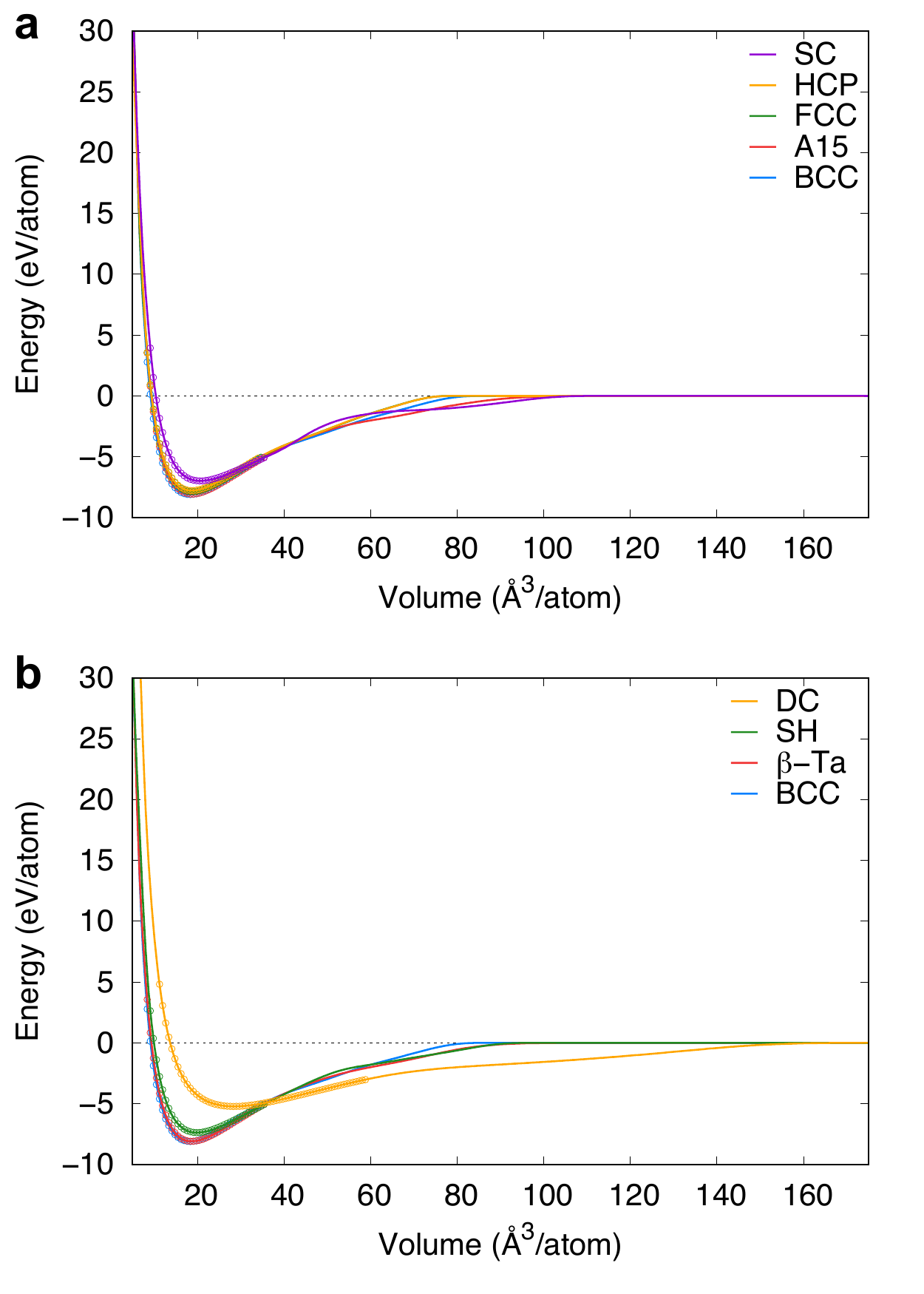}
\par\end{centering}
\caption{\label{fig:EOS_wide}Equations of state of crystal structures predicted
by the PINN potential (lines) in comparison with the DFT calculations
(points), with the atomic volume ranging from a state of large compression
to the PINN potential cutoff. \textbf{a} Simple cubic (SC), hexagonal
close-packed (HCP), face-centered cubic (FCC), A15 (Cr$_{3}$Si prototype),
and body-centered cubic (BCC) structures. \textbf{b} Diamond cubic
(DC), simple hexagonal (SH), $\beta$-Ta, and BCC structures.}
\end{figure}

\begin{figure}
\noindent \begin{centering}
\includegraphics{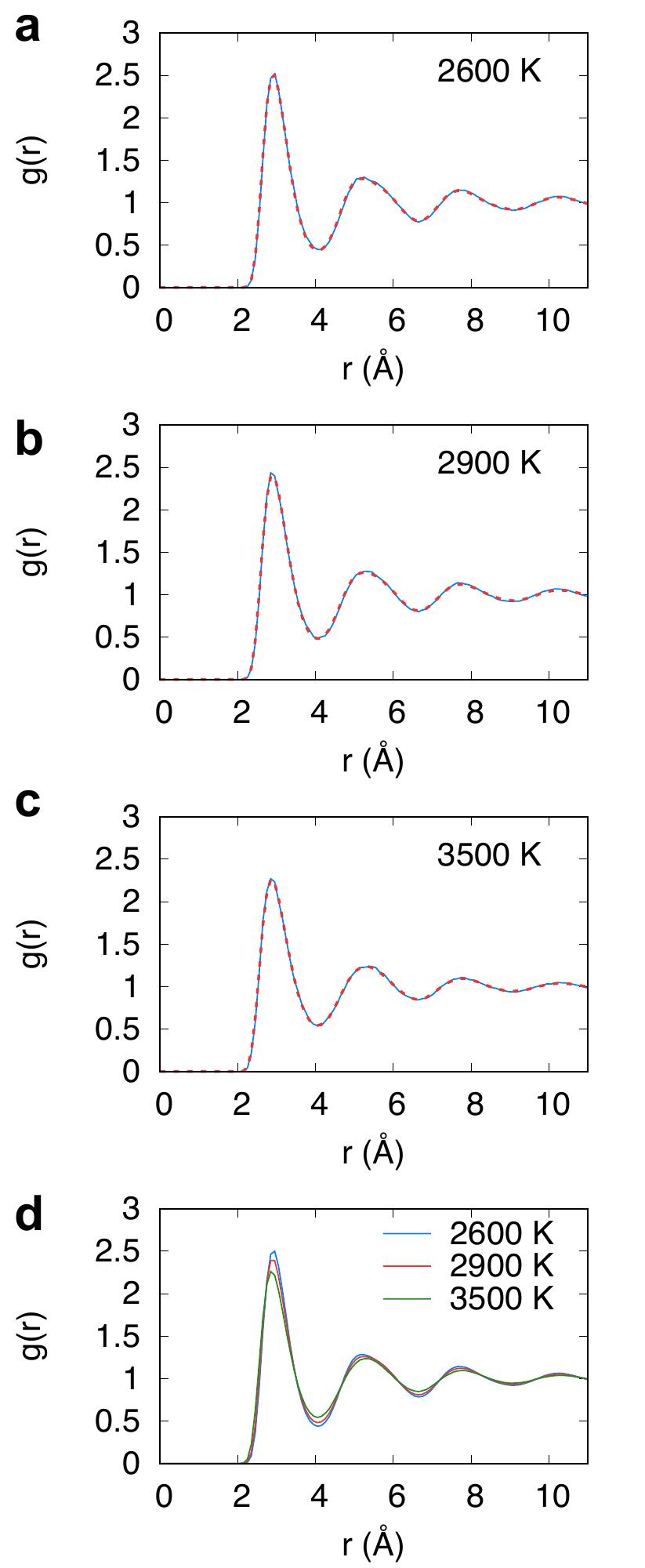}
\par\end{centering}
\caption{\textbf{\label{fig:RDF-1}a--c} Radial distribution function (RDF)
predicted by the PINN potential (red dashed lines) in comparison with
DFT calculations (blue lines) at three temperatures. \textbf{d} Variation
of the RDF with temperature predicted by the PINN potential.}
\end{figure}

\begin{figure}
\noindent \begin{centering}
\includegraphics{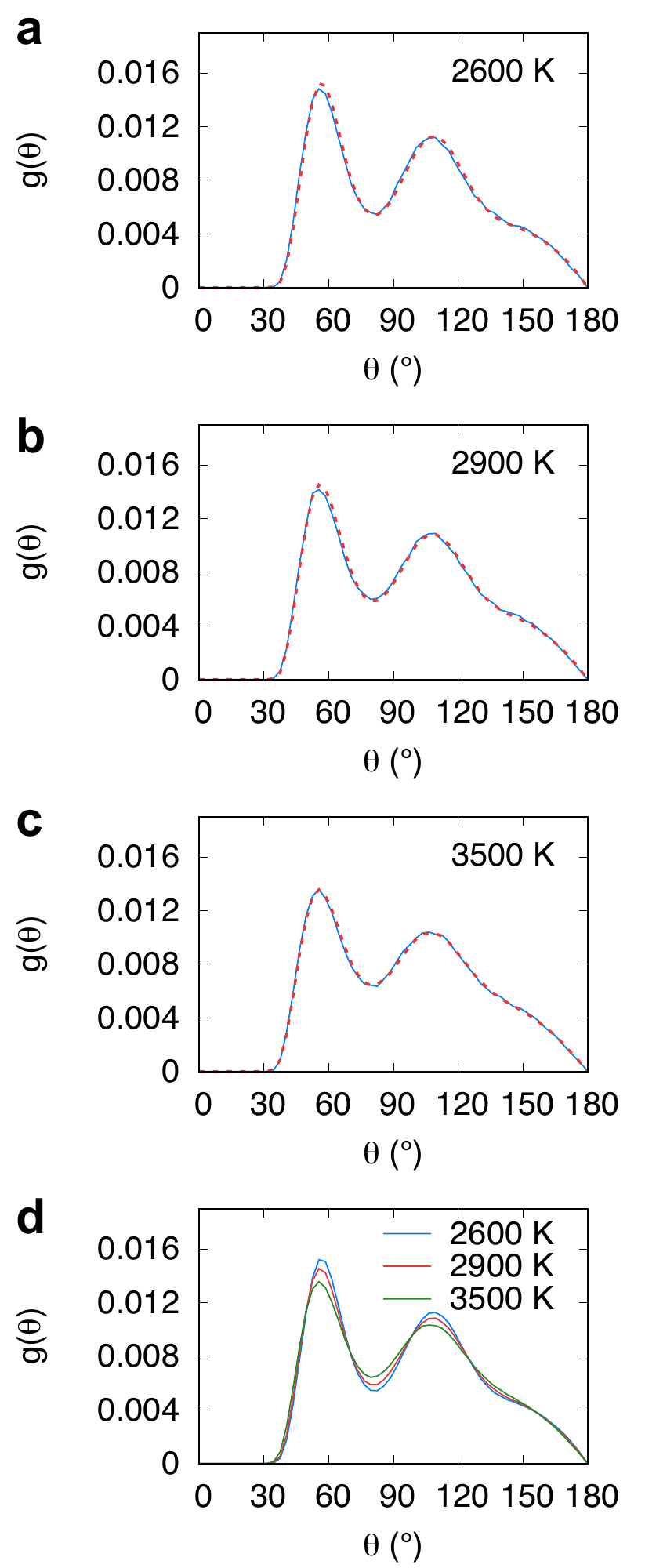}
\par\end{centering}
\caption{\textbf{\label{fig: BondAD}a--c} Bond angle distribution (BAD) predicted
by the PINN potential (red dashed lines) in comparison with DFT calculations
(blue lines) at three different temperatures. \textbf{d} BAD variation
with temperature predicted by the PINN potential.}
\end{figure}

\begin{figure}
\noindent \begin{centering}
\includegraphics{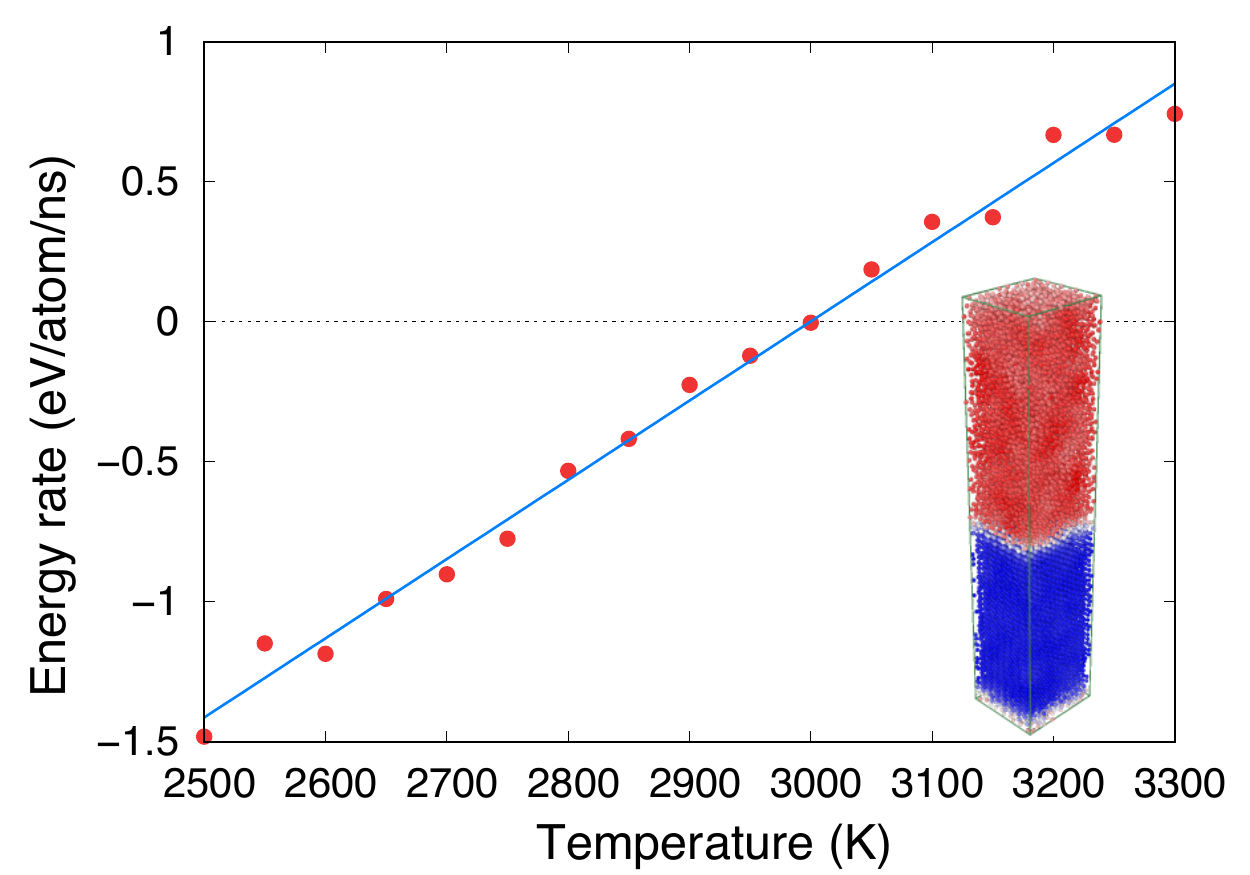}
\par\end{centering}
\caption{\label{fig:FitTm}Energy rate in MD simulations of the solid-liquid
interface computed using the PINN potential at different temperatures
(red dots). The linear fit of the data (blue line) was used to obtain
the melting temperature of Ta predicted by the potential. The inset
shows the simulation block containing the solid (blue) and liquid
(red) phases.}
\end{figure}

\begin{figure}
\noindent \begin{centering}
\includegraphics[width=1.05\columnwidth]{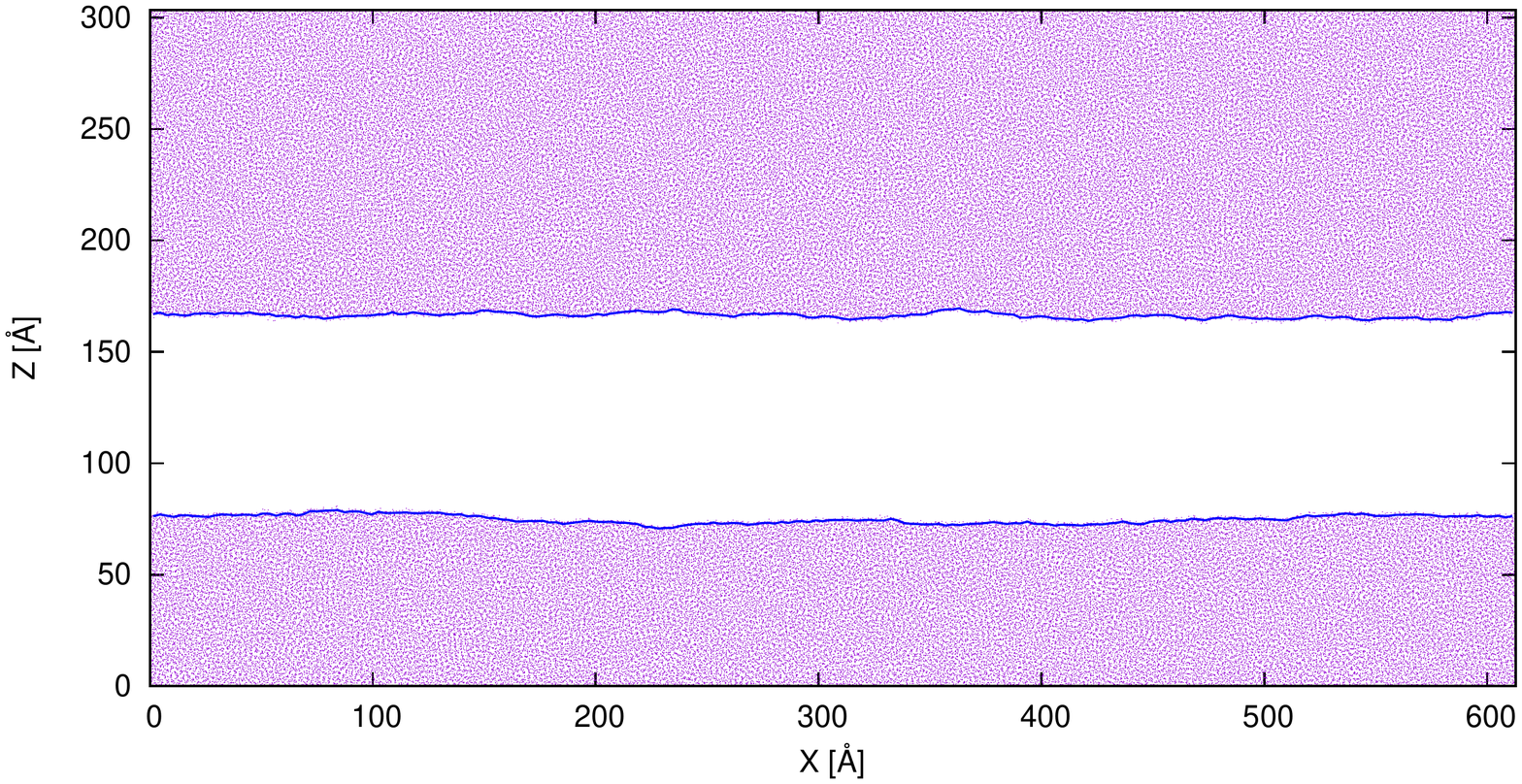}
\par\end{centering}
\caption{\label{fig:FitTm-1}Simulation block used for computing the liquid
surface tension of Ta. The liquid layers are shown in purple and the
surfaces are outlined in blue. Periodic boundary conditions are applied
in the $x$ and $y$ directions.}
\end{figure}

\begin{figure}
\noindent \begin{centering}
\includegraphics[width=0.7\columnwidth]{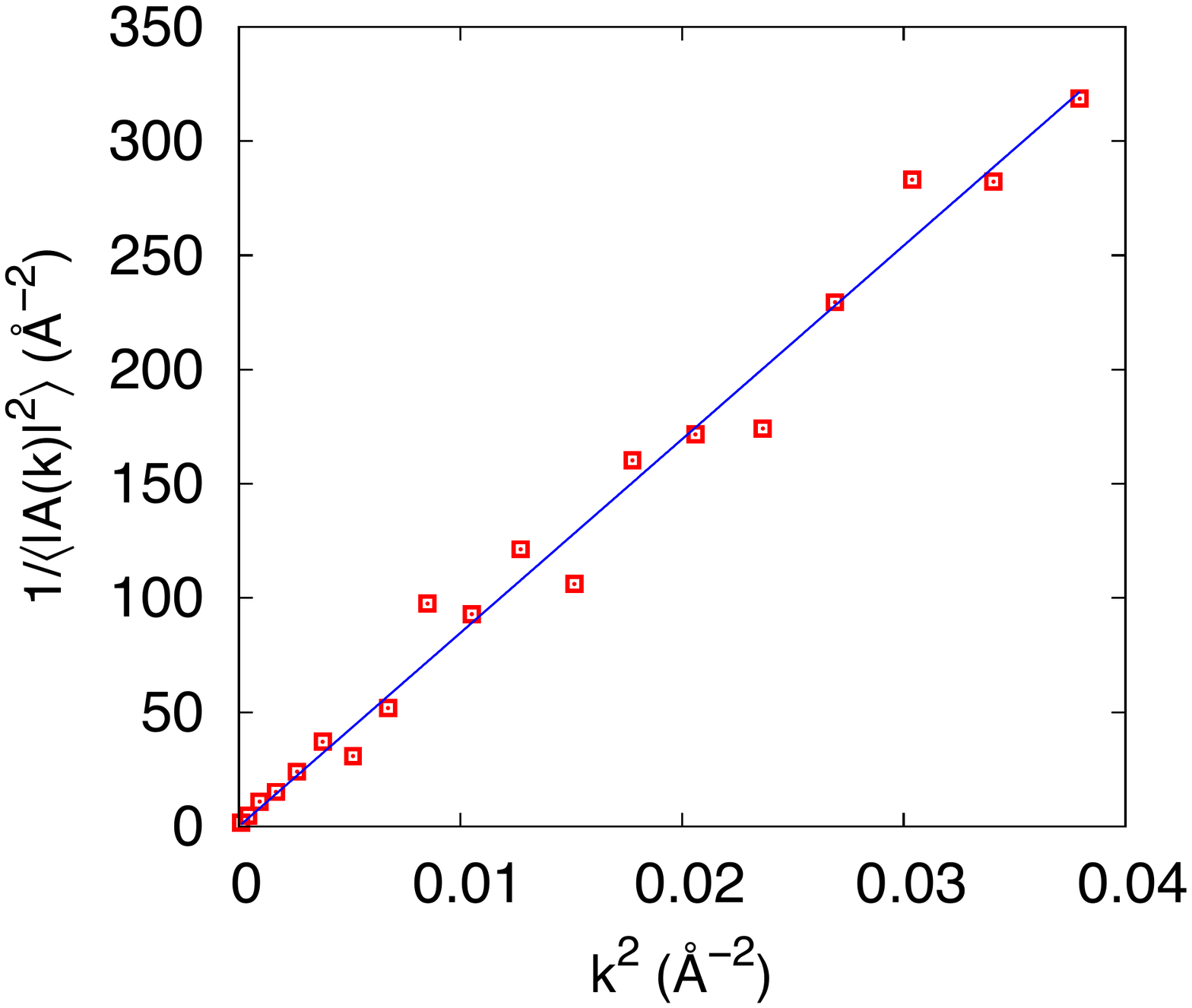}
\par\end{centering}
\caption{\label{fig:FitTm-1-1}Inverse power of capillary waves versus the
wave number squared for the liquid Ta surface computed with the PINN
potential. The line represents a linear fit in the long-wave limit.}
\end{figure}

\end{document}